\documentclass[%
reprint,
superscriptaddress,
 amsmath,
 amssymb,
 aps,
prl,
floatfix,
]{revtex4-2}

\usepackage{float}
\usepackage{braket}
\usepackage{graphicx}% Include figure files
\usepackage{dcolumn}% Align table columns on decimal point
\usepackage{bm}% bold math
\usepackage{inputenc}
\usepackage{blindtext}
\usepackage{color}
\usepackage{booktabs,siunitx}
\usepackage{comment}
\usepackage{todonotes}
\usepackage{hyperref}% add hypertext capabilities
\hypersetup{colorlinks,citecolor=blue}

\usepackage[normalem]{ulem}

\graphicspath{{Figures/}}
\newcolumntype{~}{>{\global\let\currentrowstyle\relax}}
\newcolumntype{^}{>{\currentrowstyle}}

\DeclareMathAlphabet\mathbfcal{OMS}{cmsy}{b}{n}

\begin{document}

%% commandes perso
\newcommand{\JPcomment} [1] 
{\todo[inline,backgroundcolor=green,size=\small ,bordercolor=white]{{\bf JP:} #1}}
\newcommand{\ACcomment} [1] 
{\todo[inline,backgroundcolor=cyan,size=\small ,bordercolor=white]{{\bf Alo\"{i}s:} #1}}

\newcommand{\liou}{\mathcal{L}}
\newcommand{\proj}{\mathcal{P}}
\newcommand{\qroj}{\mathcal{Q}}
\newcommand{\eigvec}{\ensuremath{\pmb{\epsilon}}}
\newcommand{\Zpart}{\ensuremath{\mathcal{Z}}}
\newcommand{\Vol}{\ensuremath{\mathcal{V}}}
\newcommand{\Cv}{\ensuremath{\mathcal{C}}}
\newcommand{\kBT}{\ensuremath{\mathrm{k}_\mathrm{B}\mathrm{T}}}
\newcommand{\Ra}{\ensuremath{\mathbfcal{R}}}
\newcommand{\bbraket}[1]{\braket{\braket{#1}}}
\newcommand{\bbl}{\ensuremath{\big(}}
\newcommand{\bbr}{\ensuremath{\big)}}
\newcommand{\KLD}{\ensuremath{D_{\mathrm{KL}}}}
\renewcommand\vec{\mathbf}
\newcommand{\phivec}{\boldsymbol{\Phi}}
\newcommand{\kcf}[2]{\big(#1, #2\big)}

\newcommand{\refcolor}{lightgray}
\newcommand {\corrections}[1]{#1}
\newcommand {\toremove}[1]{}

\preprint{APS/123-QED}

% \title{Energy conservation and the fluctuation-dissipation theorem in interacting phonon systems}
\title{Fluctuation-dissipation and virtual processes in interacting phonon systems}

%\author{}

\author{Alo\"is Castellano}
\author{J. P. Alvarinhas Batista}
\affiliation{Nanomat group, Q-MAT research center and European Theoretical Spectroscopy Facility, Université de Liège, allée du 6 août, 19, B-4000 Liège, Belgium}

\author{Matthieu J. Verstraete}
\affiliation{Nanomat group, Q-MAT research center and European Theoretical Spectroscopy Facility, Université de Liège, allée du 6 août, 19, B-4000 Liège, Belgium}
\affiliation{ITP, Physics Department, Utrecht University 3508 TA Utrecht, The Netherlands}

%\affiliation{%
% This line break forced with \textbackslash\textbackslash
%}
% 

%\date{\today}% It is always \today, today,
             %  but any date may be explicitly specified

\begin{abstract}
Phonon-phonon interactions are fundamental to understanding a wide range of material properties, including thermal transport, expansion and vibrational spectra. 
In all of the conventional perturbative approaches, energy conservation is enforced using Dirac delta functions at each microscopic phonon interaction. 
We demonstrate below that these delta functions stem from an incomplete treatment of the interacting problem, that violates the fluctuation-dissipation theorem governing systems at equilibrium. 
By replacing the deltas with convolutions of the phonon spectral functions, and introducing a self-consistency condition, we provide a more accurate and physically consistent framework. 
For systems where phonon dynamics can be approximated as Markovian, we simplify this approach, reducing the dissipative component to a single parameter tied to phonon lifetimes.
Applying this method to boron arsenide, we find that self-consistent linewidths capture the phonon scattering processes better, significantly improving agreement with experimental thermal conductivity values. 
These results also challenge the conventional view of four-phonon processes as dominant in BAs, demonstrating a three-phonon description can explain most of the experimental value, provided it is self-consistent. 
With this method we address critical limitations of perturbative approaches, offering new insights into dissipation and phonon-mediated processes, and enabling more accurate modeling of anharmonic materials.
\end{abstract}

%\keywords{Suggested keywords}%Use showkeys class option if keyword
                              %display desired
\maketitle

Phonons are quantized vibrational modes of a crystal lattice, and central to our understanding of a wide range of material properties.
From mechanical stability and sound propagation to optical responses and thermal transport, phonons play a key role in determining the behavior of solids, and enable many functionalities and applications.
As such, an accurate description of phonon dynamics is essential to advance our understanding of both fundamental physics and material functionality.
In the harmonic approximation, phonons are treated as non-interacting quasiparticles with infinite lifetimes.
This can naturally not be a complete description: in real systems, anharmonicity introduces phonon-phonon interactions, which lead to a broadening of phonon spectral lineshapes.
These interactions are crucial to describe key material properties quantitatively and sometimes even qualitatively.
For example, without interactions the phonon thermal conductivity is infinite, an inconsistency which is resolved with the introduction of anharmonicity.

In conventional perturbative approaches, phonon-phonon interactions are treated with strict energy conservation, enforced by Dirac delta functions of the energy difference before and after each scattering event~\cite{Maradudin1962, Li2012, Li2014, Han2022}.
In this letter, we show that while this formalism captures the idealized dynamics of non-interacting phonons, it employs a strong approximation that neglects the dissipative component of real phonon dynamics, and breaks the fluctuation-dissipation theorem.
We show that a more exact formulation replaces delta functions with spectral function convolutions and a self-consistency condition, which can be interpreted as high-order effective interactions mediated by virtual phonons. Importantly, these are not subject to energy conservation.
A practical implementation is introduced for cases where phonon dynamics are approximated as Markovian, reducing the dissipative component to a single parameter associated with phonon lifetimes.

We are interested in phonon correlation functions at finite temperature, which are linked to observables such as X-ray or neutron scattering.
For the derivation, we place ourselves within our recently introduced mode-coupling theory of anharmonic lattice dynamics~\cite{Castellano2023,Castellano2025,supp}, but similar arguments will hold for all of the commonly used (perturbative) theories.
Within this formalism, the phonon Kubo correlation functions (KCF)~\cite{Kubo1966} obey the generalized Langevin equation of motion
\begin{equation}
\label{eq:motion}
    \ddot{G}_\lambda(t) = -\Omega_\lambda^2 G_\lambda(t) - \int_0^t ds K_\lambda(s) \dot{G}_\lambda(t-s),
\end{equation}
where we introduce the super-index $\lambda=(\vec{q}, s)$, with $\vec{q}$ the wave vector in reciprocal space and $s$ the index of the phonon branch.
This equation describes the dynamics as driven by a harmonic force with the spring magnitude given by the phonon frequency $\Omega_\lambda$, and a time-dependent friction term. The latter is contained in a memory kernel $K_\lambda(t)$ which describes the dissipation of the phonon energy through a bath made up of all other phonons.
Importantly, $K_\lambda(t)$ must respect the fluctuation-dissipation theorem~\cite{Mori1965,Zwanzig1961,Kubo1966}, which states that $K_\lambda(t) = \big(\delta f_\lambda, \delta f_\lambda(t)\big)$, where $\delta f_\lambda(t)$ is the ``random" force which describes the effects of neglected degrees of freedom on the dynamics.
The dynamical susceptibility (or spectral function) is obtained through a Laplace transform of eq.(\ref{eq:motion}), giving
\begin{equation}
\label{eq:spectral function}
    \chi_\lambda''(\omega) = \frac{1}{\hbar\pi} \frac{4\omega \Omega_\lambda \Gamma_\lambda^2(\omega)}{(\omega^2 - \Omega_\lambda^2 - 2 \omega \Delta_\lambda(\omega))^2 + 4 \omega^2 \Gamma_\lambda^2(\omega)},
\end{equation}
where $\Gamma_\lambda(\omega)$ and $\Delta_\lambda(\omega)$ are respectively the real and imaginary parts of its memory kernel (which are related by a Kramers-Kronig transform).
The main complication of eq.(\ref{eq:spectral function}) lies in the expression of the memory kernel.
Within the mode-coupling approximation, $K$ is obtained by projecting the component of the ``random" force onto multiple displacements, introducing interactions between phonons.
To simplify our presentation, we will stay at the lowest order of phonon-phonon interaction, but it is important to note that our reasoning holds identically with higher orders.
With only three-phonon interactions and after a decoupling of four-point correlations into products of two-point correlations~\cite{supp}, the real part of the memory kernel has the form
\begin{equation}
\label{eq:memory kernel lowest order}
    \Gamma_\lambda(\omega) = \frac{1}{16} \sum_{\lambda'\lambda''} \vert \Psi_{\lambda\lambda'\lambda''} \vert^2 S_{\lambda'\lambda''}(\omega).
\end{equation}
In this equation, $\Psi_{\lambda\lambda'\lambda''}$ are third-order scattering matrix elements, and $S$ is a function describing the effect of phonons $\lambda'$ and $\lambda''$ on the memory kernel. $S$ is expressed as
\begin{equation}
    S_{\lambda'\lambda''}(\omega) = \frac{\hbar\Omega_\lambda}{n(\omega)\hbar\omega}\int_{-\infty}^{\infty} d\omega' G_{\lambda'}^<(\omega')G_{\lambda''}^<(\omega-\omega'),
\end{equation}
where the lesser phonon correlation function $G_\lambda^<(\omega)$ is proportional to the KCF, and related to the dynamical susceptibility by the fluctuation-dissipation theorem, through $G_\lambda^<(\omega) = n(\omega) \chi_\lambda''(\omega)$. $n = (e^{\beta\hbar\omega} - 1)^{-1}$ is the Bose-Einstein distribution.

A common simplification involves approximating  $S_{\lambda', \lambda''}(\omega)$ using the memory-less dynamical susceptibility of the phonons  $\lambda'$ and $\lambda''$.
This results in  $S_{\lambda', \lambda''}(\omega)$ being expressed as a sum of delta functions centered at sums or differences of phonon frequencies, weighted by phonon occupation numbers~\cite{supp}
\begin{equation}
\label{eq:s delta}
\begin{split}
    S_{\lambda'\lambda''}(\omega) &= (n_{\lambda'} + n_{\lambda''} + 1) \delta(\omega - \Omega_{\lambda'} - \Omega_{\lambda''}) \\
     &- (n_{\lambda'} + n_{\lambda''} + 1) \delta(\omega + \Omega_{\lambda'} + \Omega_{\lambda''}) \\
    &+ (n_{\lambda'} - n_{\lambda''}) \delta(\omega + \Omega_{\lambda'} - \Omega_{\lambda''}) \\
    &- (n_{\lambda'} - n_{\lambda''}) \delta(\omega - \Omega_{\lambda'} + \Omega_{\lambda''}),
\end{split}
\end{equation}
where $n_{\lambda}=n(\Omega_{\lambda})$. The delta functions in this equation have a clear physical interpretation: they represent energy conservation during each phonon interaction.
However, this approach neglects a critical element: the memory kernel.
In a memory-less framework, phonons are idealized as infinitely long-lived quasiparticles, with their dynamical susceptibility given by $\chi_{\lambda}''(\omega) \propto \delta(\omega - \Omega_\lambda)$.
This idealization inherently assumes non-dissipative dynamics, so that energy conservation arises automatically from the system’s behavior.
Introducing the memory kernel fundamentally changes this picture. 
Phonons are no longer viewed as perfectly conserving energy through their dynamics; instead, the memory kernel accounts for the exchange of energy between interacting phonons. 
In this framework, the energy conservation for individual events is generalized to the detailed balance condition, which gives rise to the fluctuation-dissipation theorem. 
Energy conservation should only be found between the initial and final states.
Consequently, the delta functions, which cannot capture dissipation, are replaced by a convolution term in eq.(\ref{eq:memory kernel lowest order}), which is self-consistently linked to eq.(\ref{eq:spectral function}).
This convolution reflects the interplay between phonons and their self-created environment, where the memory kernel describes the “phonon bath” formed by the collective dynamics of all phonons.
Modifying the memory kernel alters the phonon properties, which in turn reshape the bath—a feedback mechanism captured by the self-consistent framework.

\begin{figure}[tb]
    \centering
    \includegraphics[width=\linewidth]{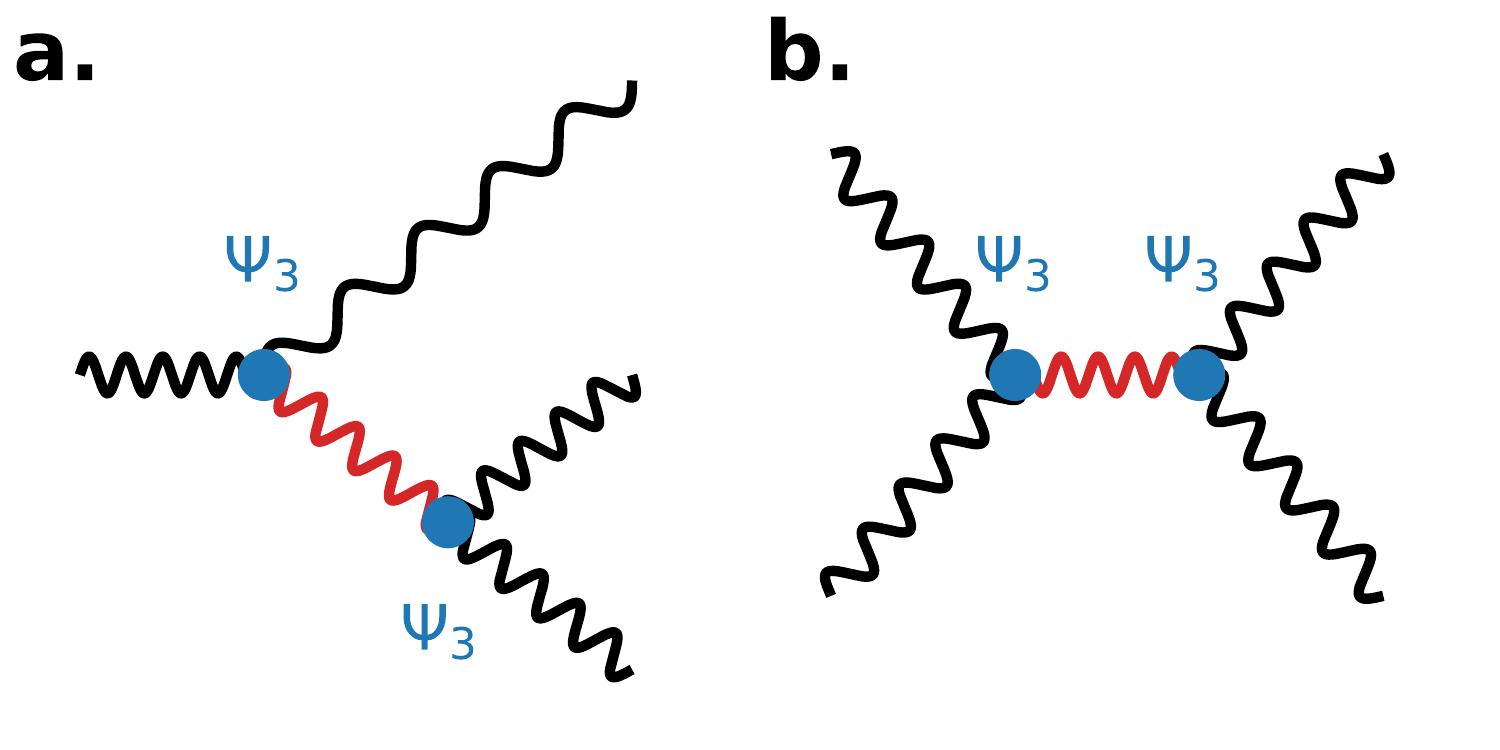}
    \caption{Examples of second-order three-phonon Feynman diagrams for anharmonic systems. The black wiggly lines represent phonon propagators, the red ones representing intermediate virtual phonons and the blue dots represent 3-phonon vertices.
    \textbf{a.} Second order decay process.
    \textbf{b.} Second order redistribution process.}
    \label{fig:feynman}
\end{figure}
In a diagrammatic framework, the convolution can be interpreted as introducing higher-order interactions mediated by virtual phonons~\cite{Carruthers1962,Leggett1965}.
For instance, the process illustrated in FIG.\ref{fig:feynman}\textbf{a.} represents an effective four-phonon interaction, where one phonon decays into three phonons through an intermediate process involving a virtual phonon. 
Similarly, FIG.\ref{fig:feynman}\textbf{b.} depicts a second-order redistribution process in which two phonons merge to form a virtual phonon, which subsequently decays into two new phonons.
In such higher-order interactions, energy and momentum conservation are strictly enforced only for the initial and final phonons.
The virtual phonons, however, are not subject to these constraints, enabling phonon interactions that would otherwise be forbidden by strict energy conservation.
When self-consistency is applied, even more complex higher-order interactions emerge, and at full convergence, an infinite series of such interactions contributes to the phonon linewidths.
Consequently, these previously forbidden interactions, which are simply a manifestation of the fluctuation-dissipation condition, can have drastic consequences on the properties of systems where selection rules could restrict anharmonicity~\cite{Lin2024,Lindsay2015,Ravichandran2020,Ravichandran2021,Yang2021}.
For instance, in systems with large mass differences - and thus a large acoustic-optical band gap - interactions between acoustic and optical modes are no longer forbidden within a three-phonon truncation.
This should modify substantially the scattering phase space and reduce, for example, the expected large conductivity in these systems.

The self consistency of the full frequency dependent dynamical susceptibility represents an important computational challenge due to the convolution appearing in $S_{\lambda',\lambda''}(\omega)$.
However, phonons are often accurately described using a Markovian approximation, where the memory kernel reduces to a single number $\Gamma_\lambda = \Gamma_\lambda(\Omega_\lambda)$ and $\chi_\lambda''(\omega)$ to a sum of Lorentzian centered on $\pm\Omega_\lambda$ with widths $\Gamma_\lambda$.
In this case, the $S_{\lambda',\lambda''}(\omega)$ function has the simple form~\cite{supp}
\begin{equation}
\label{eq:s lorentzian}
\begin{split}
    S_{\lambda'\lambda''}(\omega) &= (n_{\lambda'} + n_{\lambda''} + 1) L(\omega, \Omega_{\lambda'} + \Omega_{\lambda''}, \Gamma_{\lambda'} + \Gamma_{\lambda''}) \\
     &- (n_{\lambda'} + n_{\lambda''} + 1) L(\omega, -\Omega_{\lambda'} - \Omega_{\lambda''}, \Gamma_{\lambda'} + \Gamma_{\lambda''}) \\
    &+ (n_{\lambda'} - n_{\lambda''}) L(\omega, \Omega_{\lambda'} - \Omega_{\lambda''}, \Gamma_{\lambda'} + \Gamma_{\lambda''}) \\
    &- (n_{\lambda'} - n_{\lambda''})L(\omega, -\Omega_{\lambda'} + \Omega_{\lambda''}, \Gamma_{\lambda'} + \Gamma_{\lambda''}),
\end{split}
\end{equation}
where $L$ are Lorentzian functions defined as $L(\omega, \Omega, \Gamma) = \Gamma / ((\omega - \Omega)^2 + \Gamma^2) / \pi$.
Thus, within the Markovian approximation, the fluctuation-dissipation theorem can be simply enforced by using this equation in a self-consistent cycle involving equations (\ref{eq:spectral function}), (\ref{eq:memory kernel lowest order}) and (\ref{eq:s lorentzian}).
This represents a simplification compared to the numerical implementation of delta functions~\cite{Yates2007,Li2012,Li2014,Han2022} as it avoids the need of introducing numerical convergence parameters and usually creates a broader smearing than would be used numerically, easing convergence. It does however bring an extra-cost related to the need to compute eq.(\ref{eq:memory kernel lowest order}) several times.
It should be noted that the need for linewidth self-consistency has already been proposed in previous work~\cite{Gu2019, Turney2009}.
However, these works do not propose a formal justification for this self-consistency.
Moreover, the smearing used in these works does not correspond to the convolution of the Lorentzian dynamical susceptibility appearing in eq.(\ref{eq:s lorentzian}), breaking the relation between the dissipative component of the phonon dynamics and the memory kernel, and with it, the fluctuation-dissipation theorem.

\begin{figure}
    \centering
    \includegraphics[width=\linewidth]{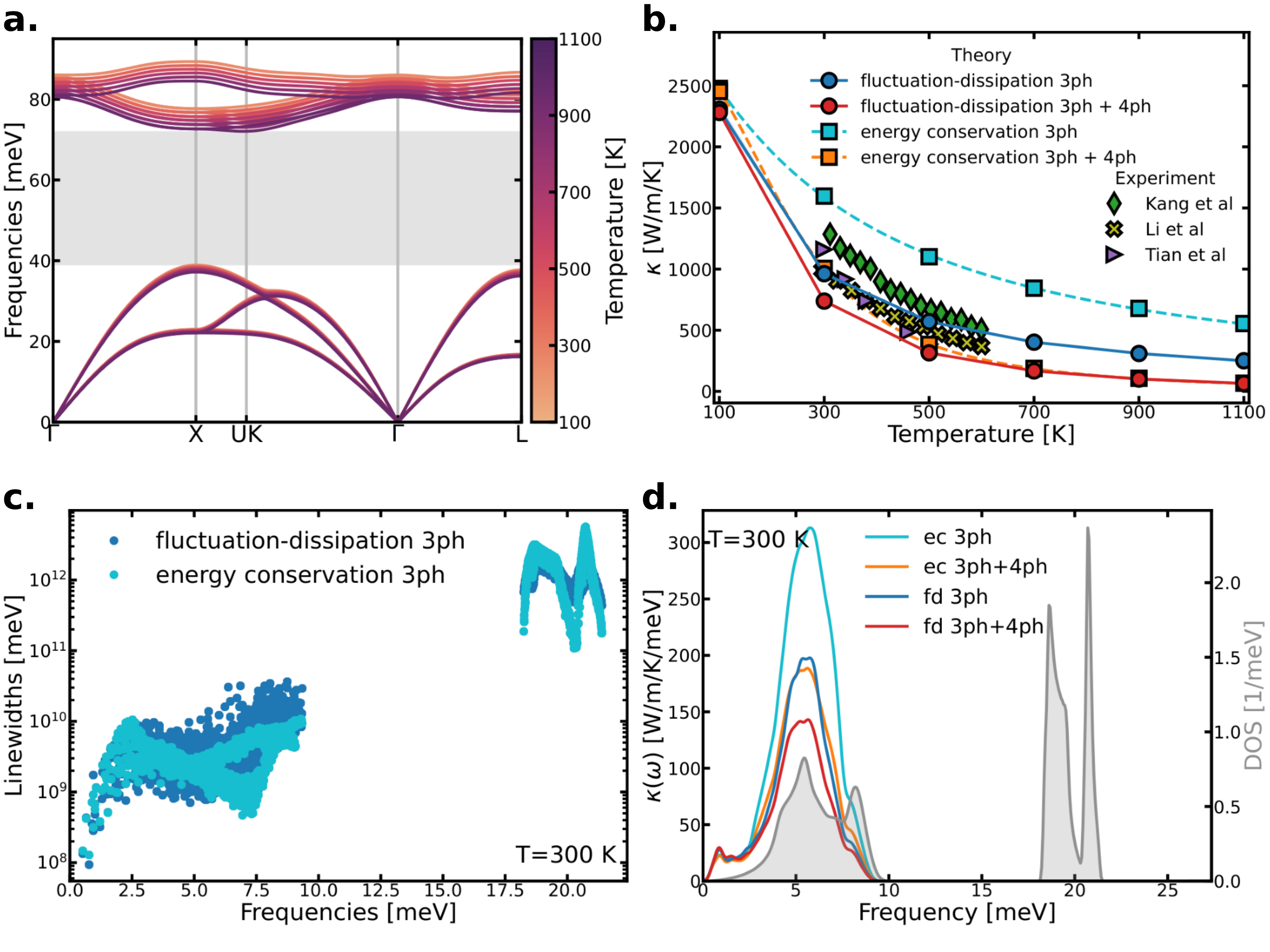}
    \caption{\textbf{a.} T-dependent phonons in BAs from 100 to 1100 K. The large acoustic to optical band-gap is highlighted in grey.
    \textbf{b.} Thermal conductivity in BAs including 3 or 3+4 phonon interactions, with energy conservation enforced or using the fluctuation-dissipation condition. Experimental results from Kang \textit{et al}~\cite{Kang2018}, Li \textit{et al}~\cite{Li2018} and Tian \textit{et al}~\cite{Tian2018} are shown for comparison.
    \textbf{c.} Comparison of the phonon linewidths at $300$~K with 3-phonon interactions with energy conservation vs fluctuation-dissipation condition.
    \textbf{d.} Comparison of the spectral thermal conductivity at $300$~K computed with energy conservation (ec) or after self-consistency to enforce the fluctuation-dissipation condition (fd).
    The phonon density of states is represented in grey.}
    \label{fig:results}
\end{figure}

To demonstrate the importance of the fluctuation-dissipation condition, we applied our approach to boron arsenide, a semiconductor renowned for its exceptionally high thermal conductivity~\cite{Kang2018,Li2018,Tian2018,Lindsay2013}.
A key feature of this material is its large band gap separating the acoustic and optical phonon branches, as shown in FIG.\ref{fig:results}\textbf{a}.
This gap is sufficiently wide to forbid three-phonon coupling between acoustic and optical phonons if strict energy conservation is applied~\cite{Ravichandran2020}, making four-phonon interactions a requirement to accurately reproduce the thermal conductivity in this system~\cite{Feng2017}.
However, when the energy conservation principle is replaced by the fluctuation-dissipation theorem, third-order acoustic-optical interactions are no longer strictly forbidden: through the Markovian approximation
physically natural Lorentzian functions~\cite{supp} appear. Their long tails extend the scattering phase space, significantly altering the dynamics of both acoustic and optical phonons.
This highlights BAs as an archetypal system where the fluctuation-dissipation theorem should play a critical role in capturing the true phonon scattering processes, and their impact on thermal transport.

Our results in FIG.\ref{fig:results} were computed using machine-learning interatomic potentials constructed on \textit{ab initio} data, whose details are provided in the SM~\cite{supp}.
At the three-phonon level, the impact of the fluctuation-dissipation condition is particularly striking. 
While enforcing strict energy conservation allows us to reproduce theoretical results from the literature~\cite{Lindsay2013,Feng2017,Ravichandran2020}, incorporating self-consistent linewidths consistently leads to lower thermal conductivity values, which are in closer agreement with experimental measurements~\cite{Tian2018,Kang2018,Li2018}. 
This reduction can be attributed to a significant decrease in the thermal conductivity contribution from phonons with energies around $20$~meV. 
Under strict energy conservation, these phonons are forbidden from interacting with optical modes, but the convolution in the self-consistent approach relaxes this restriction, thereby expanding the scattering phase space.
This behavior is analogous to what occurs when four-phonon processes are included: the increased scattering phase space results in higher linewidths for these phonons and a corresponding reduction in their thermal conductivity contribution. 
Our findings challenge the prevailing view that four-phonon interactions are essential to explain the thermal conductivity in large phonon band gap materials like boron arsenide. 
While these interactions are not negligible, their importance appears to have been overestimated in previous literature. 
Remaining at the three-phonon level while enforcing the fluctuation-dissipation condition provides a reasonably accurate prediction of thermal transport in this material, consistent with its low anharmonicity.
\corrections{However, this accuracy should not be seen as an indication that the self-consistent determination of linewidth could avoid the need for four-phonon interactions.
Instead, the fluctuation-dissipation condition should be seen as a new, superior framework to evaluate all scattering mechanisms. Its importance in the context of BAs stems from the constrained 3-phonon scattering phase-space produced by strict energy conservation. The FD approach removes this constraint, and shows the reproduction of the experimental thermal conductivity of BAs using energy conservation and the 4-phonon channel is actually a coincidence: the EC 3-phonon $\kappa$ is too high, and the EC 4-phonon correction is too large, but happens to land near the experimental values.}

The fluctuation-dissipation condition is universally applicable, but its impact on different materials varies, and requires quantitative evaluation.
\toremove{In the SM~\cite{supp}, we demonstrate that the impact of self-consistency is negligible for silicon — a highly harmonic system with low linewidths and no phonon band gap — whereas it is significant for silver iodide, which exhibits strong anharmonicity.}
\corrections{In the SM~\cite{supp}, we show the impact of the self-consistency for silicon, a highly harmonic system with low linewidths and no phonon band gap, and for silver iodide, which exhibits strong anharmonicity.
In the former case, the difference in using the fluctuation-dissipation condition is negligible, both with three and four-phonon interactions.
In AgI however, its impact is more pronounced.
Including only three-phonon scattering, the self-consistency reduces the thermal conductivity by 15\% to 25\% between 100 to 300~K.
When four phonon scattering is included, the impact of the fluctuation-dissipation condition is smaller, between 5\% and 20~\% in the same range of temperature, which is still significant for low thermal conductivity materials.}
\toremove{Ultimately, we anticipate that self-consistency will substantially alter the phonon dynamics in systems with large mass differences or pronounced anharmonicity. 
For example, hydrogen-rich materials, which feature both characteristics, are promising candidates for further exploration of the fluctuation-dissipation condition’s role in phonon interactions.}
\corrections{Ultimately, the fluctuation-dissipation condition is a universal principle that applies to all materials, yet its impact on phonon dynamics appears to be highly system-dependent.
Our findings highlight the need for broader investigations to determine how this condition interacts with anharmonicity and the available scattering phase space.
In more complex or strongly interacting systems, additional factors, such as possible deviations from Markovian behavior, may also influence the outcome.
Such studies will be essential for establishing when and how fluctuation-dissipation-consistent treatments are necessary for accurate modeling of lattice dynamics.}

While our study focuses on phonons, similar self-consistent approaches should be relevant for other finite-temperature quasiparticles. 
Systems involving interactions between quasiparticles with vastly different energy scales, such as magnon-phonon or electron-phonon interactions, will be particularly affected by the relaxation of strict energy conservation. 
For instance, recent work~\cite{Lihm2024,Lihm2024b} demonstrates that incorporating self-consistent electronic linewidths resolves divergences previously expected in piezoelectric materials. 
This highlights the broader importance of capturing dissipative dynamics in a rigorous and consistent framework, extending the relevance of our approach beyond phonons to other complex quasiparticle systems.

\begin{acknowledgments}
We thank S. Poncé and J-M. Lihm for discussions about self-consistent electron phonon calculations.
The authors acknowledge the Fonds de la Recherche Scientifique (FRS-FNRS Belgium) and Fonds Wetenschappelijk Onderzoek (FWO Belgium) for EOS project CONNECT (G.A. 40007563), and 
F\'ed\'eration Wallonie Bruxelles and ULiege for funding ARC project DREAMS (G.A. 21/25-11).
MJV acknowledges funding by the Dutch Gravitation program
“Materials for the Quantum Age” (QuMat, reg number 024.005.006), financed by the Dutch Ministry of Education, Culture and Science (OCW).

% CPU
Simulation time was awarded by 
by PRACE on Discoverer at SofiaTech in Bulgaria (optospin project id. 2020225411); 
EuroHPC-JU award EHPC-EXT-2023E02-050 on MareNostrum 5 at Barcelona Supercomputing Center (BSC), Spain;
by the CECI (FRS-FNRS Belgium Grant No. 2.5020.11);
and by the Lucia Tier-1 of the F\'ed\'eration Wallonie-Bruxelles (Walloon Region grant agreement No. 1117545).
\end{acknowledgments}

\bibliography{biblio}% Produces the bibliography via BibTeX.

\end{document}

%% commandes perso
\newcommand{\JPcomment} [1] 
{\todo[inline,backgroundcolor=green,size=\small ,bordercolor=white]{{\bf JP:} #1}}
\newcommand{\ACcomment} [1] 
{\todo[inline,backgroundcolor=cyan,size=\small ,bordercolor=white]{{\bf Alo\"{i}s:} #1}}

\newcommand{\liou}{\mathcal{L}}
\newcommand{\proj}{\mathcal{P}}
\newcommand{\qroj}{\mathcal{Q}}
\newcommand{\eigvec}{\ensuremath{\pmb{\epsilon}}}
\newcommand{\Zpart}{\ensuremath{\mathcal{Z}}}
\newcommand{\Vol}{\ensuremath{\mathcal{V}}}
\newcommand{\Cv}{\ensuremath{\mathcal{C}}}
\newcommand{\kBT}{\ensuremath{\mathrm{k}_\mathrm{B}\mathrm{T}}}
\newcommand{\Ra}{\ensuremath{\mathbfcal{R}}}
\newcommand{\bbraket}[1]{\braket{\braket{#1}}}
\newcommand{\bbl}{\ensuremath{\big(}}
\newcommand{\bbr}{\ensuremath{\big)}}
\newcommand{\KLD}{\ensuremath{D_{\mathrm{KL}}}}
\renewcommand\vec{\mathbf}
\newcommand{\phivec}{\boldsymbol{\Phi}}
\newcommand{\kcf}[2]{\big(#1, #2\big)}

\newcommand{\refcolor}{lightgray}
\newcommand{\corrections}[1]{#1}
\newcommand {\toremove}[1]{}

\renewcommand{\thesection}{S\arabic{section}}
\renewcommand{\theequation}{S\arabic{equation}}
\renewcommand{\thefigure}{S\arabic{figure}}
\renewcommand{\bibnumfmt}[1]{[S#1]}
\renewcommand{\citenumfont}[1]{S#1}

\preprint{APS/123-QED}

\title{Supplementary material: Fluctuation-dissipation and virtual processes in interacting phonon systems}

\author{Alo\"is Castellano}
\author{J. P. Alvarinhas Batista}
\affiliation{Nanomat group, Q-MAT research center and European Theoretical Spectroscopy Facility, Université de Liège, allée du 6 août, 19, B-4000 Liège, Belgium}

\author{Matthieu J. Verstraete}
\affiliation{Nanomat group, Q-MAT research center and European Theoretical Spectroscopy Facility, Université de Liège, allée du 6 août, 19, B-4000 Liège, Belgium}
\affiliation{ITP, Physics Department, Utrecht University 3508 TA Utrecht, The Netherlands}

\maketitle

\section{The mode-coupling theory of anharmonic lattice dynamics}

In this section, we describe briefly the mode-coupling theory of anharmonic lattice dynamics, and we refer the interested reader to references~\cite{Castellano2023} and \cite{Castellano2025} for more details.
For our purposes, we introduce the Kubo correlation function, defined as
\begin{equation}
\label{eq:Kubo corr}
    \kcf{A}{B(t)} = \kBT \int_0^\beta d\lambda \braket{A(i\hbar\lambda)B(t)}.
\end{equation}
We assume that atoms vibrate around equilibrium positions $\braket{\vec{R}}$ without diffusion, allowing the introduction of mass-weighted displacements $u_i = \sqrt{M_i}(R_i - \braket{R_i})$, where $i$ is a composite index for an atom $I$ in a Cartesian direction $\alpha$, $\braket{R_i}$ its average position and $M_i$ is the mass of the corresponding atom.
In these mass-scaled coordinates, the forces are expressed with $f_i = -\nabla_{\vec{R}_i} V(\vec{R}) / \sqrt{M_i}$.
The aim of the theory is to describe the displacement-displacement Kubo correlation function
\begin{equation}
\label{eq:displ Kubo corr}
    G_{ij} = \kcf{u_i}{u_j(t)}.
\end{equation}
The first step in the derivation is built on the Mori-Zwanzig projection operator formalism, in which the projection operator $\proj$ and its orthogonal projection $\qroj$ are defined as
\begin{align}
\label{eq:projectors}
    \proj =& \sum_{ij} \bigg[\frac{\kcf{u_j}{\mathcal{O}}}{\kcf{u_i}{u_j}}u_i + \frac{\kcf{p_j}{\mathcal{O}}}{\kcf{p_i}{p_j}}p_i\bigg], \\
    \qroj =& 1 - \proj.
\end{align}
Applying the identity $\proj + \qroj = 1$ on the forces separates out the ``known'' and ``random'' parts. Taking the Kubo correlation of the result with respect to displacements, one obtains an equation of motion for the displacement-displacement correlation function as the generalized Langevin equation
\begin{equation}
    \ddot{G}_{ij}(t) = -\sum_{k}\Phi_{jk} G_{ik}(t) - \sum_{k}\int_0^t ds K_{jk}(s) \dot{G}_{ik}(t-s).
\end{equation}
In this equation, we introduced the frequency matrix
\begin{equation}
\label{eq:frequency matrix}
    \Phi_{ij} = \sum_k \kcf{u_k}{u_j}^{-1}\kcf{u_k}{f_i}
\end{equation}
as well as the memory kernel,
\begin{align}
\label{eq:memory kernel}
    K_{ij}(t) = \frac{1}{\kBT} \kcf{\delta f_i}{\delta f_j(t)},
\end{align}
which is the Kubo correlation function of the random forces
\begin{equation}
\label{eq:random forces}
    \delta f_j(t) = e^{i\qroj\liou t} \bigg( f_j + \sum_k \Phi_{jk} u_k\bigg).
\end{equation}
The generalized interatomic force constants can be Fourier transformed and diagonalized (note that the masses are already included in $\boldsymbol{\Phi}$ through the mass-scaled coordinates):
\begin{equation}
    \boldsymbol{\Phi}(\vec{q}) \boldsymbol{\varepsilon}_\lambda = \Omega_\lambda^2 \boldsymbol{\varepsilon}_\lambda,
\end{equation}
allowing the introduction of phonons with frequency $\Omega_\lambda$ and eigenvectors $\boldsymbol{\varepsilon}_\lambda$, where $\lambda=(\vec{q}, s)$ is a super-index with $\vec{q}$ the wavevector and $s$ a phonon branch.
Consequently, the dynamics of the system can be described in terms of phonon operators $A_\lambda$ and random forces $\delta A_\lambda$, defined with the projection
\begin{align}
    A_\lambda(t) =& \sqrt{\frac{2 \Omega_\lambda}{\hbar}} \sum_i \varepsilon_\lambda^i u_i(t) \\
    \delta A_\lambda(t) =& \sqrt{\frac{2 \Omega_\lambda}{\hbar}} \sum_i \varepsilon_\lambda^i \delta f_i(t).
\end{align}
The memory kernel in phonon coordinates becomes
\begin{equation}
    K_\lambda(t) = \frac{\hbar}{2 \kBT \Omega_\lambda} \big(\delta A_\lambda, \delta A_\lambda(t) \big),
\end{equation}
allowing to introduce the generalized Langevin equation for the motion of a phonon $\lambda$
\begin{equation}
\label{eq:motion phonons}
    \ddot{G}_{\lambda}(t) = -\Omega_\lambda^2 G_\lambda(t) - \int_0^t ds K_\lambda(s) \dot{G}_\lambda(t-s).
\end{equation}

The most complicated part of this equation is the expression of the memory kernel.
In the mode-couping theory, the difficulties are alleviated by introducing higher-order projection on several displacements.
To lowest order, the random forces in real space are approximated as
\begin{equation}
\begin{split}
    \delta f_i \approx& -\frac{1}{2}\sum_{jklm} \frac{\big(u_l u_m, \delta f_i \big)}{\big(u_j u_k, u_l u_m\big)} u_j u_k \\
    =& -\frac{1}{2}\sum_{jk} \Psi_{ijk} u_j u_k,
\end{split}
\end{equation}
where $\Psi_{ijk}$ are generalized third-order force constants.
Projecting on phonons, the random forces are expressed as
\begin{equation}
    \delta A_\lambda(t) \approx \sqrt{\frac{\hbar}{2}} \frac{\Omega_\lambda}{2} \sum_{\lambda'\lambda''} \Psi_{\lambda\lambda'\lambda''} A_{\lambda'}(t) A_{\lambda''}(t),
\end{equation}
with the three-phonon scattering matrix elements
\begin{equation}
\begin{split}
    \Psi_{\lambda\lambda'\lambda''} =& \frac{1}{\sqrt{\Omega_\lambda \Omega_{\lambda'} \Omega_{\lambda''}}} \sum_{ijk} \frac{\varepsilon_\lambda^i \varepsilon_{\lambda'}^j \varepsilon_{\lambda''}^k }{\sqrt{M_i M_j M_k}} \Psi_{ijk} \\
    \times& e^{i(\vec{R}_i \vec{q}_\lambda + \vec{R}_j \vec{q}_{\lambda'} + \vec{R}_k \vec{q}_{\lambda''})} \Delta_{\vec{G}}(\vec{q} + \vec{q}' + \vec{q}'').
\end{split}
\end{equation}
The resulting memory kernel for a phonon $\lambda$ is finally written
\begin{equation}
\label{eq:memory kernel phonons}
\begin{split}
    K_\lambda(t) =& \frac{1}{16} \frac{\hbar \Omega_\lambda}{\kBT} \sum_{\lambda'\lambda''\lambda'''\lambda''''} \Psi_{\lambda\lambda'\lambda''}\Psi_{\lambda\lambda'''\lambda''''} \\
    &\big( A_{\lambda'} A_{\lambda''}, A_{\lambda'''}(t) A_{\lambda''''}(t) \big).
\end{split}
\end{equation}

\section{The S function}

To express the memory kernel, one needs to compute the four-point correlation of eq.(\ref{eq:memory kernel phonons}).
Carefully accounting for the imaginary time integral of Kubo correlations, we perform a decoupling~\cite{Castellano2025} expressed through a Fourier transform as:
\begin{equation}
\begin{split}
    &\int dt \big(A_{\lambda_1} A_{\lambda_2}, A_{\lambda_3}(t) A_{\lambda_4}(t) \big) e^{i\omega t} \\
    &\approx\frac{\kBT}{n(\omega)\hbar \omega} \int_{-\infty}^{\infty} d\omega' G_{\lambda_1}^<(\omega') G_{\lambda_2}^<(\omega - \omega') \\ 
    &\times (\delta_{\lambda_1 \lambda_3}\delta_{\lambda_2 \lambda_4} + \delta_{\lambda_1 \lambda_4}\delta_{\lambda_2 \lambda_3}),
\end{split}
\end{equation}
where
\begin{equation}
    n(\omega) = \frac{1}{e^{\beta \hbar\omega} -1}
\end{equation}
is the Bose-Einstein distribution function.
Multiplying this result by $\hbar\Omega_\lambda / \kBT$, we can now introduce the $S$ function of the main text as
\begin{equation}
\label{eq:S function full}
S_{\lambda'\lambda''}(\omega) = \frac{\hbar\Omega_\lambda}{n(\omega)\hbar\omega} \int_{-\infty}^{\infty} d\omega' G_{\lambda'}^<(\omega') G_{\lambda''}^<(\omega - \omega'),
\end{equation}
where we also introduced the lesser correlation function
\begin{equation}
    G_{\lambda}^<(\omega) = n(\omega) \chi_\lambda''(\omega)
\end{equation}
Injecting eq.(\ref{eq:S function full}) into eq.(\ref{eq:memory kernel phonons}), the memory kernel is approximated as
\begin{equation}
\label{eq:memory kernel S function}
    \Gamma_\lambda(\omega) \approx \frac{1}{16} \sum_{\lambda' \lambda''} \vert \Psi_{\lambda\lambda'\lambda''}\vert^2 S_{\lambda'\lambda''}(\omega).
\end{equation}

With the presence of Kubo correlation functions in the definition of the $S$ function, eq.(\ref{eq:memory kernel S function}) has to be solved in a self-consistent cycle involving eq.(\ref{eq:motion phonons})

\subsection{Memory-less approximation}

As suggested by its name, the memory-less approximation consists in neglecting the phonon dissipation in their equation of motion, which becomes
\begin{equation}
    \ddot{G}_\lambda(t) = -\Omega_\lambda^2 G_\lambda(t).
\end{equation}
Such an approximation corresponds to the usual non-interacting phonons, with a spectral function given by
\begin{equation}
\label{eq:chi memory-less}
    \chi_{\lambda}''(\omega) = \frac{\omega}{4\hbar\Omega_\lambda} \bigg[\delta(\omega - \Omega_{\lambda}) + \delta(\omega + \Omega_{\lambda}) \bigg].
\end{equation}
Introducing this equation into the $S$ function gives rise to convolutions of the form
\begin{equation}
\begin{split}
    &\int_{-\infty}^{\infty} d\omega' n(\omega') \delta(\omega \pm \Omega_{\lambda'}) n(\omega - \omega') \delta(\omega - \omega' \pm \Omega_{\lambda''}) \\
    &= n(\mp \Omega_{\lambda'}) n(\mp \Omega_{\lambda''}) \delta(\omega \pm \Omega_{\lambda'} \pm \Omega_{\lambda''}).
\end{split}
\end{equation}
For further simplification, we also use the identity of the Bose-Einstein distribution for sums of frequencies
\begin{equation}
\label{eq:Bose-Einstein sum}
    n(\Omega_{\lambda'} + \Omega_{\lambda''}) = \frac{n_{\lambda'} n_{\lambda''}}{(n_{\lambda'} + 1)(n_{\lambda''}+1) - n_{\lambda'} n_{\lambda''}},
\end{equation}
where we introduced the notation $n_\lambda = n(\Omega_\lambda)$.
Injecting all these results into the $S$ function, we obtain finally
\begin{equation}
\label{eq:S energy conservation}
\begin{split}
    S_{\lambda'\lambda''}(\omega) =& (n_{\lambda'} - n_{\lambda''}) \delta(\omega + \Omega_{\lambda'} - \Omega_{\lambda''}) \\
    -& (n_{\lambda'} - n_{\lambda''}) \delta(\omega - \Omega_{\lambda'} + \Omega_{\lambda''}) \\
    +& (n_{\lambda'} + n_{\lambda''} + 1) \delta(\omega - \Omega_{\lambda'} - \Omega_{\lambda''}) \\
    -& (n_{\lambda'} + n_{\lambda''} + 1) \delta(\omega + \Omega_{\lambda'} + \Omega_{\lambda''}),
\end{split}
\end{equation}
which is the usual result for phonon-phonon interactions with microscopic energy conservation.

An important thing to notice about this derivation is that the energy conservation is a direct consequence of the neglect of the memory-kernel and is consequently only exact for non-interacting phonons, hence in a purely (eventually effective) harmonic system.

\subsection{Markovian approximation}

The Markovian approximation is obtained by assuming that the phonons are fast compared to the random forces, and consists in taking the $t\rightarrow \infty$ limit in the convolution term of eq.(\ref{eq:motion phonons}).
In the mode-coupling approximation, this results in a Langevin equation of motion for the phonon Kubo correlation function
\begin{equation}
    \ddot{G}_\lambda(t) = -\Omega_\lambda^2 G_\lambda(t) - \Gamma_\lambda \dot{G}_{\lambda}(t),
\end{equation}
where $\Gamma_\lambda = \Gamma_\lambda(\Omega_\lambda)$.
In frequency space, the Markovian dynamical susceptibility has a Lorentzian shape:
\begin{equation}
\begin{split}
    \chi_{\lambda}''(\omega) =& \frac{1}{\pi} \frac{4\omega \Omega_\lambda \Gamma_\lambda}{(\omega^2 - \Omega_\lambda^2)^2 + 4 \omega^2 \Gamma_\lambda^2} \\
    \approx& \frac{\omega}{\Omega_\lambda\pi} \bigg[\frac{\Gamma_\lambda}{(\omega - \Omega_\lambda)^2 + \Gamma_\lambda^2} + \frac{\Gamma_\lambda}{(\omega + \Omega_\lambda)^2 + \Gamma_\lambda^2}\bigg].
\end{split}
\label{eq:MarkovSusceptibility}
\end{equation}
Introducing the following notation for the Lorentzian
\begin{equation}
    L(\omega, \Omega, \Gamma) = \frac{1}{\pi} \frac{\Gamma}{(\omega - \Omega)^2 + \Gamma^2}
\end{equation}
and assuming a small linewidth $\Gamma_\lambda$ compared to the frequency $\Omega_\lambda$, the lesser phonon correlation functions can be approximated as
\begin{equation}
    G_{\lambda}^<(\omega) \approx n(\Omega_\lambda) L(\omega, \Omega_\lambda, \Gamma_\lambda) - n(-\Omega_\lambda) L(\omega, -\Omega_\lambda, \Gamma_\lambda).
\end{equation}
Thus, the $S$ function will be composed of convolutions of the form
\begin{equation}
\begin{split}
    &n(\pm\Omega_{\lambda'}) n(\pm\Omega_{\lambda''}) \bigg(L(\cdot, \pm \Omega_{\lambda'}, \Gamma_{\lambda'}) \ast L(\cdot, \pm \Omega_{\lambda''}, \Gamma_{\lambda''})\bigg)(\omega) \\
    &\approx \frac{n(\mp \Omega_{\lambda'}) n(\mp \Omega_{\lambda''})}{n(\omega)} L(\omega, \pm \Omega_{\lambda'} \pm \Omega_{\lambda''}, \Gamma_{\lambda'} + \Gamma_{\lambda''}).
\end{split}
\end{equation}
Assuming a small enough linewidth, so that the Lorentzian in this equation can be approximated as acting as a delta function, the prefactor of can be rewritten $n(\pm\Omega_{\lambda'}) n(\pm\Omega_{\lambda''}) / n(\pm\Omega_{\lambda'}\pm\Omega_{\lambda''})$.
Collecting these results, the $S$ function in the Markovian approximation is written
\begin{equation}
\label{eq:S Markovian}
\begin{split}
    S_{\lambda'\lambda''}(\omega) &= (n_{\lambda'} + n_{\lambda''} + 1) L(\omega, \Omega_{\lambda'} + \Omega_{\lambda''}, \Gamma_{\lambda'} + \Gamma_{\lambda''}) \\
     &- (n_{\lambda'} + n_{\lambda''} + 1) L(\omega, -\Omega_{\lambda'} - \Omega_{\lambda''}, \Gamma_{\lambda'} + \Gamma_{\lambda''}) \\
    &+ (n_{\lambda'} - n_{\lambda''}) L(\omega, \Omega_{\lambda'} - \Omega_{\lambda''}, \Gamma_{\lambda'} + \Gamma_{\lambda''}) \\
    &- (n_{\lambda'} - n_{\lambda''})L(\omega, -\Omega_{\lambda'} + \Omega_{\lambda''}, \Gamma_{\lambda'} + \Gamma_{\lambda''}).
\end{split}
\end{equation}

%Compared to the energy conservation results given by eq.(\ref{eq:S energy conservation}), two main differences are to be noticed.
%The first one is the replacement of the delta functions by Lorentzian functions.
%This is an important change as the long-tails of the Lorentzian will bring more scattering in the memory kernel compared to the energy conservation condition.
%The second differences is in the prefactor in front of these Lorentzian.
%Since the relations $\Omega_{\lambda} = \pm \Omega_{\lambda'} \pm \Omega_{\lambda''}$ are no longer enforced by the energy conservation, the simplifications introduced by eq.(\ref{eq:Bose-Einstein sum}) no longer apply.
%This can incur changes in the strength of the scattering between phonons.

The derivation so far only includes three phonon interactions.
However, it can be important to consider other phonon-phonon contributions to the memory kernel, such as the four phonon or isotopic disorder scattering.
Renaming the previously derived memory kernel as $\Gamma_\lambda^{\mathrm{3ph}}(\omega)$, the memory kernel with these contributions is given by
\begin{equation}
    \Gamma_\lambda(\omega) = \Gamma_\lambda^{\mathrm{iso}}(\omega) + \Gamma_\lambda^{\mathrm{3ph}}(\omega) + \Gamma_{\lambda}^{\mathrm{4ph}}(\omega).
\end{equation}
The isotopic contribution $\Gamma_\lambda^{\mathrm{iso}}(\omega)$, corresponding to Tamura's model~\cite{Tamura1983}, is written
\begin{equation}
    \Gamma_\lambda^{\mathrm{iso}}(\omega) = \sum_{\lambda'} g_i \vert \epsilon_\lambda^i \epsilon_{\lambda'}^i \vert^2 L(\omega, \Omega_{\lambda'}, \Gamma_{\lambda'}),
\end{equation}
with $g_i$ measuring the distribution of the isotope masses of element $i$ and computed as $g_i=\sum_n \frac{d_{i,n}}{N} \bigg( \frac{\Delta M_{i,n}}{M_{i}} \bigg)^2$.
In this equation, $N$ is the number of isotopes, $d_{i,n}$ is the concentration of isotope $n$ of element $i$, and $\Delta M_{i,n}$ is the mass difference between isotope $n$ and the average mass of the element.
The four phonon contribution $\Gamma_\lambda^{\mathrm{4ph}}$ is given by
\begin{equation}
    \Gamma_\lambda^{\mathrm{4ph}}(\omega) = \frac{\pi}{96} \sum_{\lambda'\lambda''\lambda'''} \vert \Psi_{\lambda\lambda'\lambda''\lambda'''} \vert^2 S_{\lambda'\lambda''\lambda'''}(\omega),
\end{equation}
with $\Psi_{\lambda\lambda'\lambda''\lambda'''}$ the four-phonon scattering matrix elements.
Following the same steps as for the three phonons, the $S$ function for four phonons in the Markovian approximation is given by
\begin{widetext}
\begin{equation}
\begin{split}
    S_{\lambda'\lambda''\lambda'''} =& \frac{n_{\lambda'}n_{\lambda''}n_{\lambda'''}}{n(\omega)} L(\omega, \Omega_{\lambda'} + \Omega_{\lambda''} + \Omega_{\lambda'''}, \Gamma_{\lambda'} + \Gamma_{\lambda''} + \Gamma_{\lambda'''}) \\
    +& \frac{(n_{\lambda'} + 1) n_{\lambda''}n_{\lambda'''}}{n(\omega)} L(\omega, -\Omega_{\lambda'} + \Omega_{\lambda''} + \Omega_{\lambda'''}, \Gamma_{\lambda'} + \Gamma_{\lambda''} + \Gamma_{\lambda'''}) \\
    +& \frac{n_{\lambda'} (n_{\lambda''} + 1) n_{\lambda'''}}{n(\omega)} L(\omega, \Omega_{\lambda'} - \Omega_{\lambda''} + \Omega_{\lambda'''}, \Gamma_{\lambda'} + \Gamma_{\lambda''} + \Gamma_{\lambda'''}) \\
    +& \frac{n_{\lambda'} n_{\lambda''} (n_{\lambda'''} + 1)}{n(\omega)} L(\omega, \Omega_{\lambda'} + \Omega_{\lambda''} - \Omega_{\lambda'''}, \Gamma_{\lambda'} + \Gamma_{\lambda''} + \Gamma_{\lambda'''}) \\
    +& \frac{(n_{\lambda'} + 1) (n_{\lambda''} + 1) n_{\lambda'''}}{n(\omega)} L(\omega, -\Omega_{\lambda'} - \Omega_{\lambda''} + \Omega_{\lambda'''}, \Gamma_{\lambda'} + \Gamma_{\lambda''} + \Gamma_{\lambda'''}) \\
    +& \frac{(n_{\lambda'} + 1) n_{\lambda''} (n_{\lambda'''} + 1)}{n(\omega)} L(\omega, -\Omega_{\lambda'} + \Omega_{\lambda''} - \Omega_{\lambda'''}, \Gamma_{\lambda'} + \Gamma_{\lambda''} + \Gamma_{\lambda'''}) \\
    +& \frac{n_{\lambda'} (n_{\lambda''} + 1) (n_{\lambda'''} + 1)}{n(\omega)} L(\omega, \Omega_{\lambda'} - \Omega_{\lambda''} - \Omega_{\lambda'''}, \Gamma_{\lambda'} + \Gamma_{\lambda''} + \Gamma_{\lambda'''}) \\
    +& \frac{(n_{\lambda'} + 1) (n_{\lambda''} + 1) (n_{\lambda'''} + 1)}{n(\omega)} L(\omega, -\Omega_{\lambda'} - \Omega_{\lambda''} - \Omega_{\lambda'''}, \Gamma_{\lambda'} + \Gamma_{\lambda''} + \Gamma_{\lambda'''}).
\end{split}
\end{equation}
\end{widetext}

\section{Lorentzian and energy conservation}
\begin{figure}
    \centering
    \includegraphics[width=\linewidth]{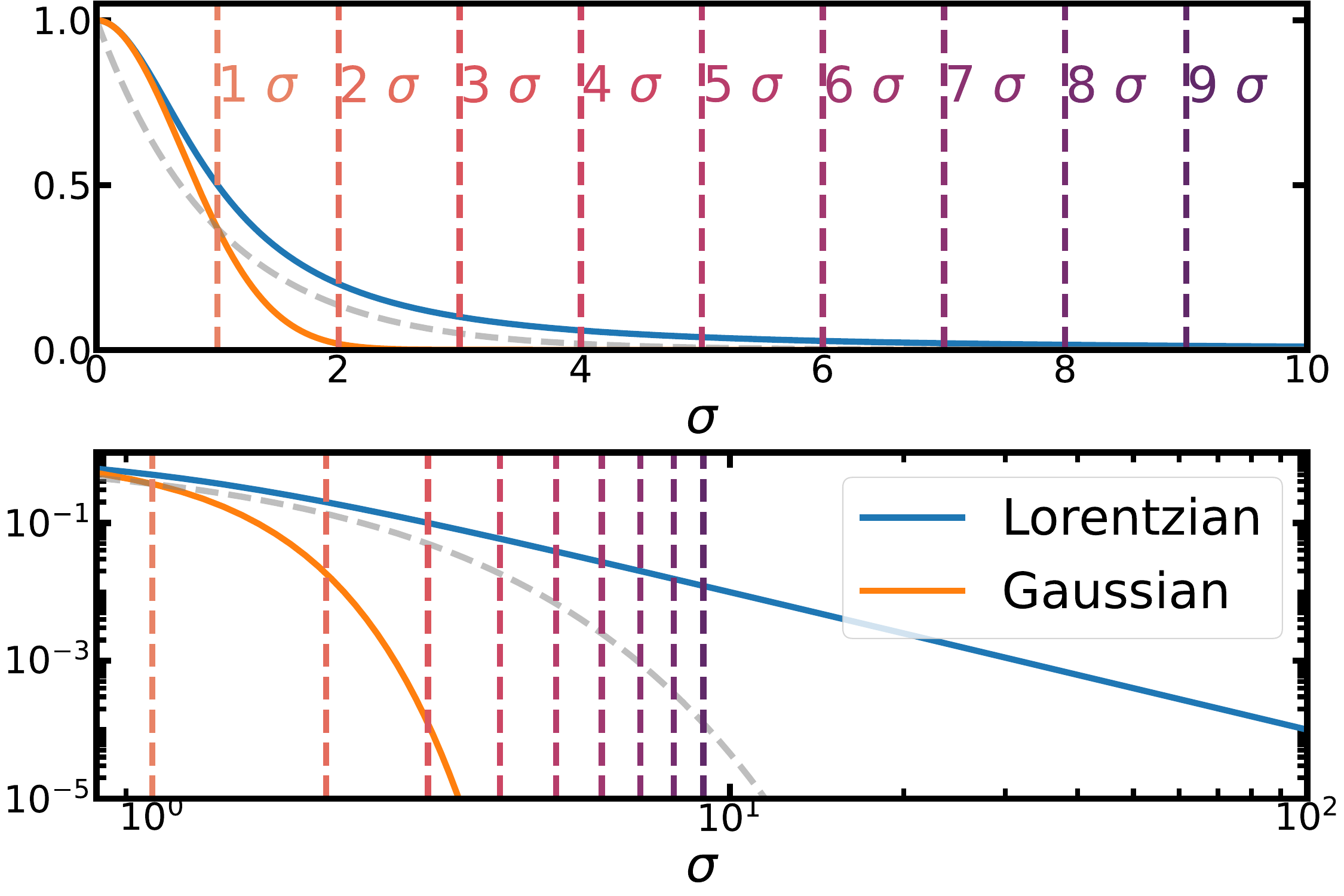}
    \caption{Lorentzian and Gaussian functions centered on 0, with equal width $\sigma=1$ and normalized to 1 at $x=0$.
    Top is in a linear scale and bottom is a log-log scale.
    The grey dashed-line shows an exponential decay.}
    \label{fig:lorgauss}
\end{figure}
\begin{figure}[thb]
    \centering
    \includegraphics[width=\linewidth]{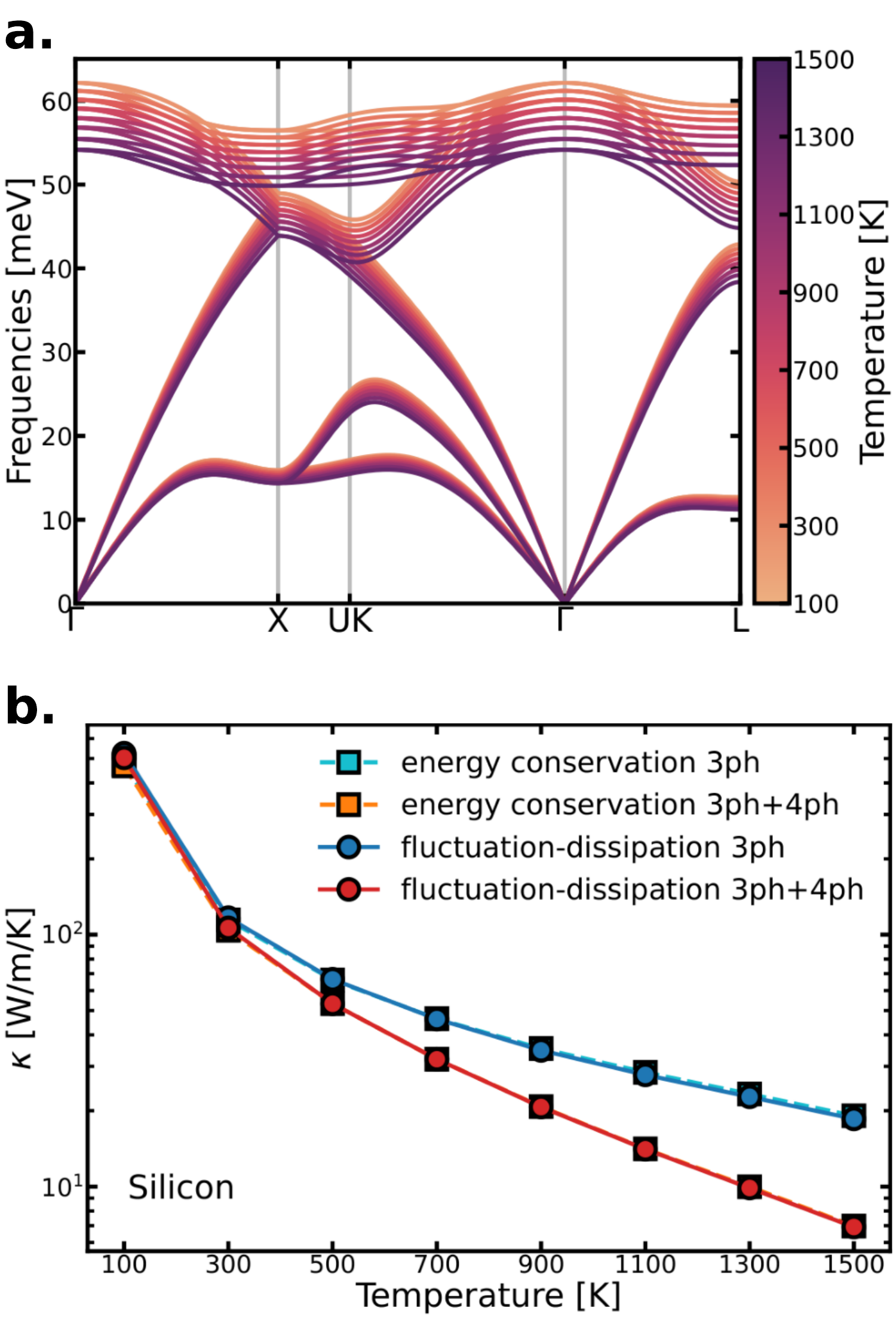}
    \caption{\textbf{a.} Evolution of the phonons in silicon between 100 and 1500~K.
    \textbf{b.} Thermal conductivity of silicon computed with energy conservation or the fluctuation-dissipation condition.}
    \label{fig:Si results}
\end{figure}
\begin{figure}[t]
    \centering
    \includegraphics[width=\linewidth]{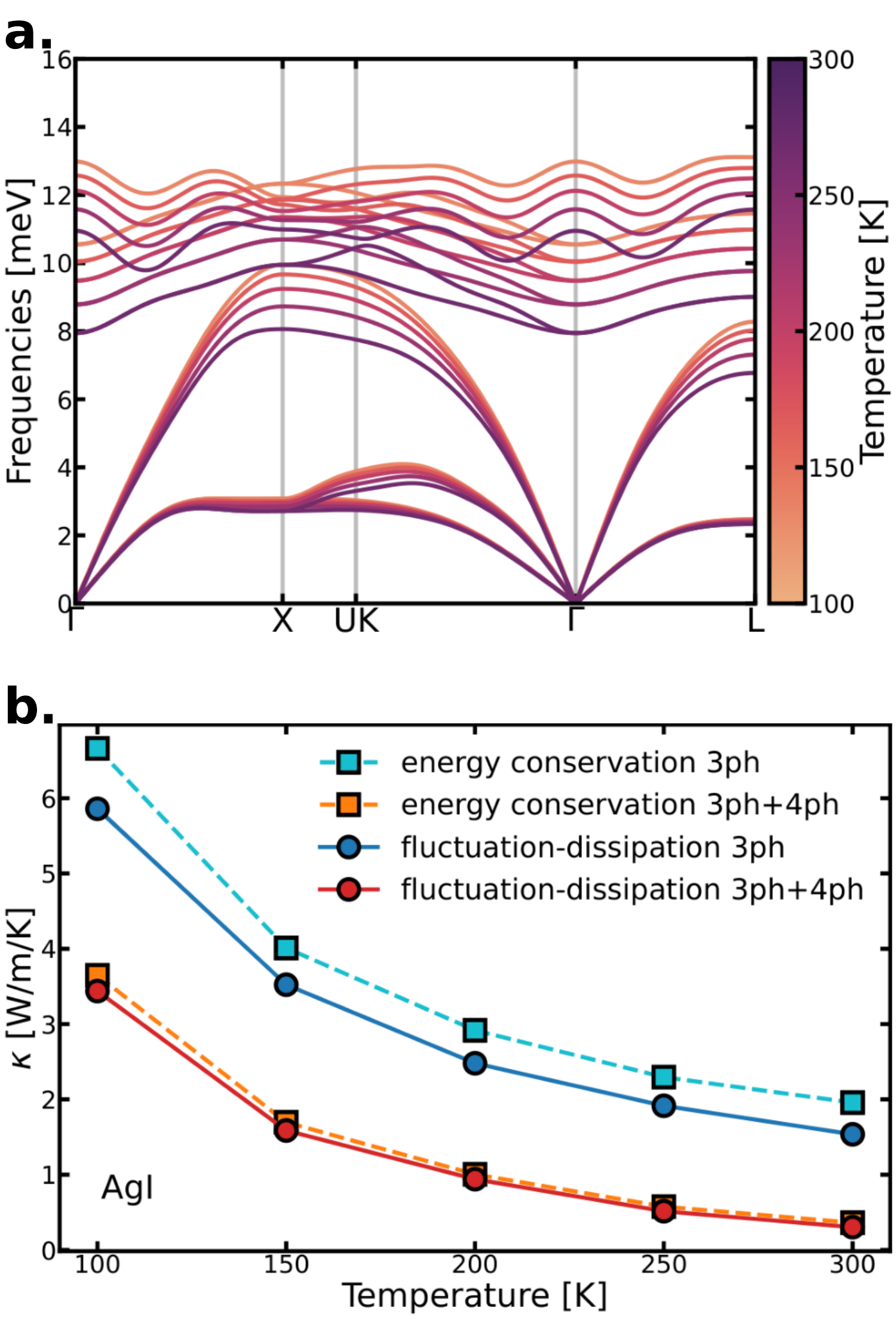}
    \caption{\textbf{a.} Evolution of the phonons in AgI between 100 and 300~K.
    \textbf{b.} Thermal conductivity of AgI computed with energy conservation or the fluctuation-dissipation condition.}
    \label{fig:AgI results}
\end{figure}

With the energy conservation condition, the numerical approximation of the delta function becomes one of the most delicate parts for the implementation of phonon-phonon interactions.
Usually, the delta are replaced by highly peaked symmetric functions which recover the delta function in the limit of vanishing width.
In recent years, the use of Gaussians with adaptively parametrized widths has become the \textit{de facto} standard, but other functions have been proposed, in particular the Lorentzian.
For highly harmonic systems, in which the linewidths are small, the fact that both Gaussian and Lorentzian are approximations of delta functions seems to imply that the energy-conservation is still somewhat valid.
The results for AgI and Si displayed on fig.\ref{fig:AgI results} and \ref{fig:Si results} seem to confirm this intuition, with $\kappa$ varying only slightly when going from the energy conservation to the fluctuation-dissipation condition.

However, the BAs result of the main text shows that the validity of energy conservation is more complex, allowing us to highlight an often overlooked but key difference between Lorentzians and Gaussians as numerical approximations for delta functions.
Indeed, if both reduce to delta peaks in the limit of vanishing width, these function have a very different behavior once their width is finite.
In particular, the Lorentzian is a long-tail distribution, decreasing exponentially slower than the Gaussian.
For instance fig.\ref{fig:lorgauss} shows that one need to go to more than $56~\sigma$ to decrease the value of a Lorentzian to $10^{-4}$.
For phonon-phonon interactions this means that, when using Lorentzians, interactions between phonons with a large energy separation still have noticeable effects on $\Gamma_\lambda$, thus breaking energy-conservation.

\begin{figure}[t]
    \centering
    \includegraphics[width=\linewidth]{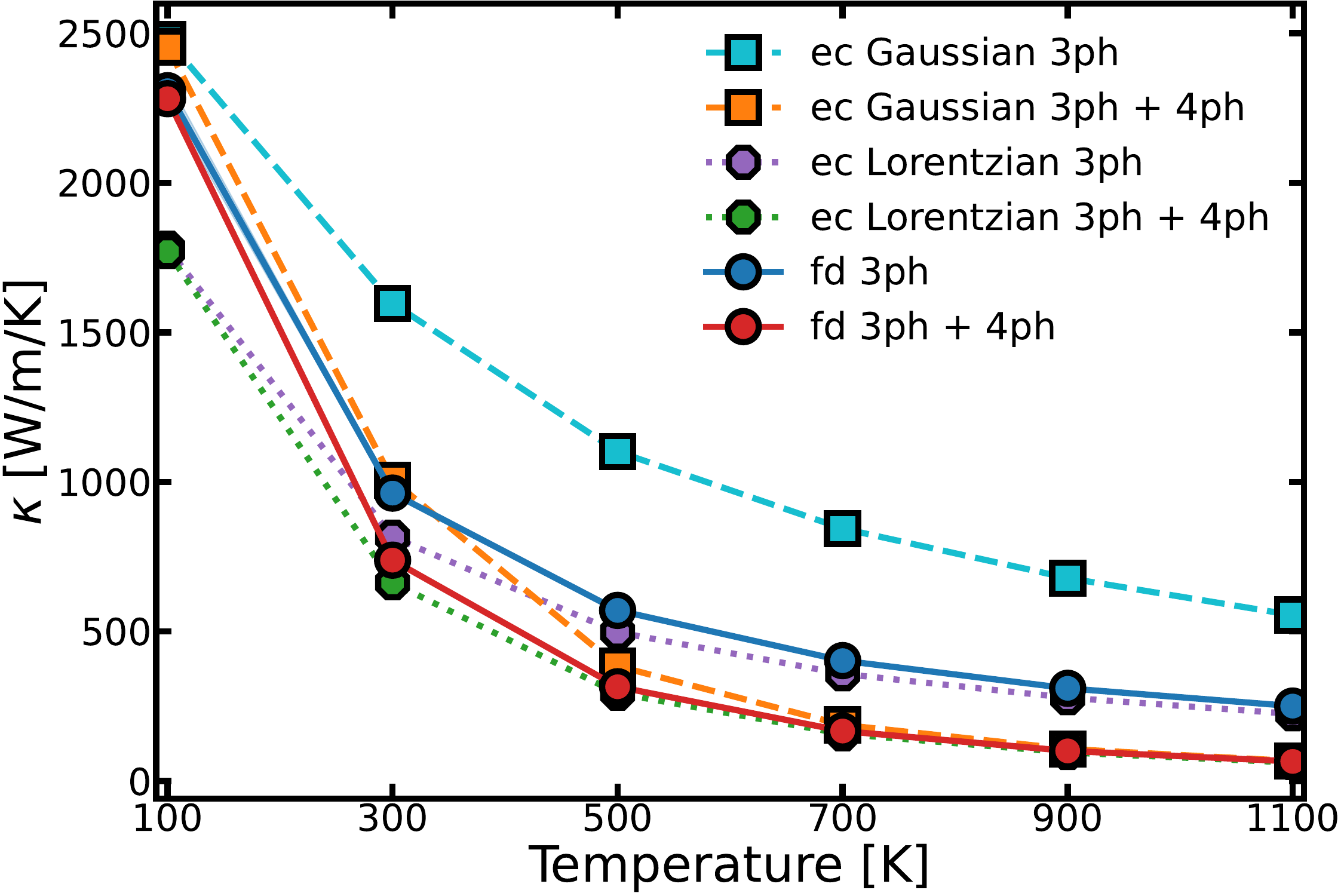}
    \caption{Thermal conductivity in BAs using energy conservation (ec), with either Gaussian or Lorentzian approximation to delta functions or with the fluctuation-dissipation condition (fd).}
    \label{fig:kappa lorentzian}
\end{figure}
This is illustrated on fig~\ref{fig:kappa lorentzian}, which shows that for BAs, enforcing energy-conservation through Lorentzian functions results in a thermal conductivity close to that obtained with the physically exact fluctuation-dissipation condition.
While this casts doubt on the Lorentzian's utility as a good numerical approximation for delta functions, it seems that the adaptive broadening scheme with Lorentzian functions can serve as an approximation for the full self-consistent calculations for the fluctuation-dissipation, especially at high T.

\begin{figure}[t]
    \centering
    \includegraphics[width=\linewidth]{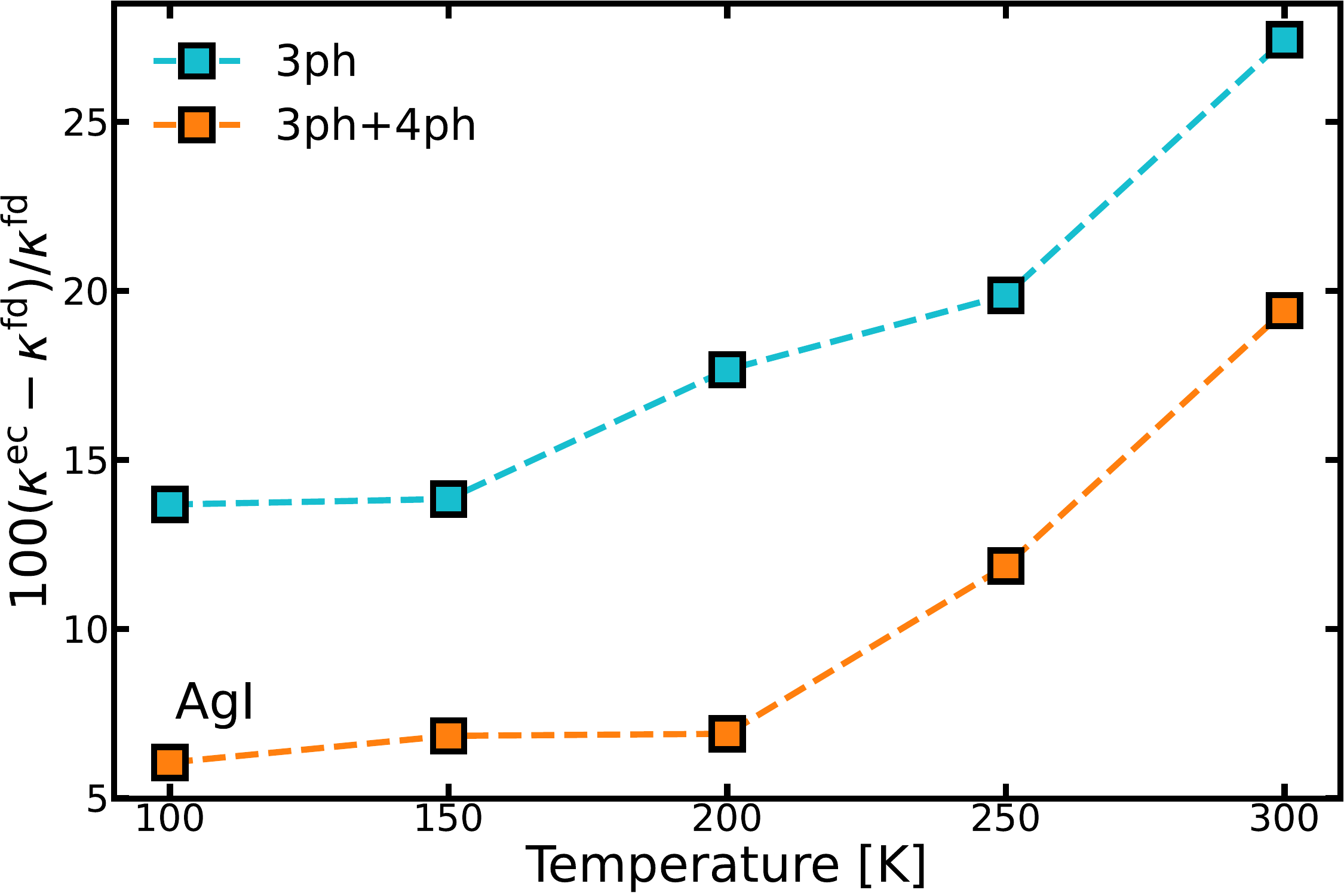}
    \caption{Difference between the thermal conductivity of AgI computed with energy conservation or the fluctuation-dissipation condition.}
    \label{fig:AgI diff}
\end{figure}

\section{Results in Silicon and AgI}
To assess the effects of fluctuation-dissipation on other systems, we compute the thermal conductivity with self-consistent linewidths for silicon and AgI, using the generalized interatomic force constants calculated in ref~\cite{Castellano2025}.

Similarly to BAs, silicon is a very harmonic material, presenting high frequencies and low linewidths, so that the phonon dynamical susceptibilities in this system are close to Lorentzian.
However, being mono-elemental, the phonon dispersion in Si doesn't present a gap between the acoustic and optical bands as shown on fig.\ref{fig:Si results}.
As a result, the thermal conductivity computed with energy conservation or the fluctuation-dissipation condition are virtually indistinguishable, as can be seen in fig.\ref{fig:Si results}.

The corresponding results for AgI are shown in fig.\ref{fig:AgI results} and \ref{fig:AgI diff}
Here anharmonicity is much stronger, as are the phonon frequency shifts with T.
\toremove{However, the very small acoustic-optical gap leads to only a modest change when going from energy conservation to the full fluctuation dissipation condition. 4th order scattering makes up more than a third of the thermal resistivity.
Once it is included the ec-fd difference is basically negligible. Comparing to BAs it is thus crucial to distinguish cases where 4th order scattering is truly essential from those where self consistency is more important.}
\corrections{Compared to BAs, the variation observed when including the fluctuation dissipation conditions are smaller, but fig.\ref{fig:AgI diff} shows that the self-consistency still amounts to a sizable reduction in the thermal conductivity
At the three-phonon level, the $\kappa$ reduction is between 15 and 25\% from 100 to 300~K, and with four-phonon interactions this reduction is still 5\% to 20\%.}

\begin{figure*}
    \centering
    \includegraphics[width=\linewidth]{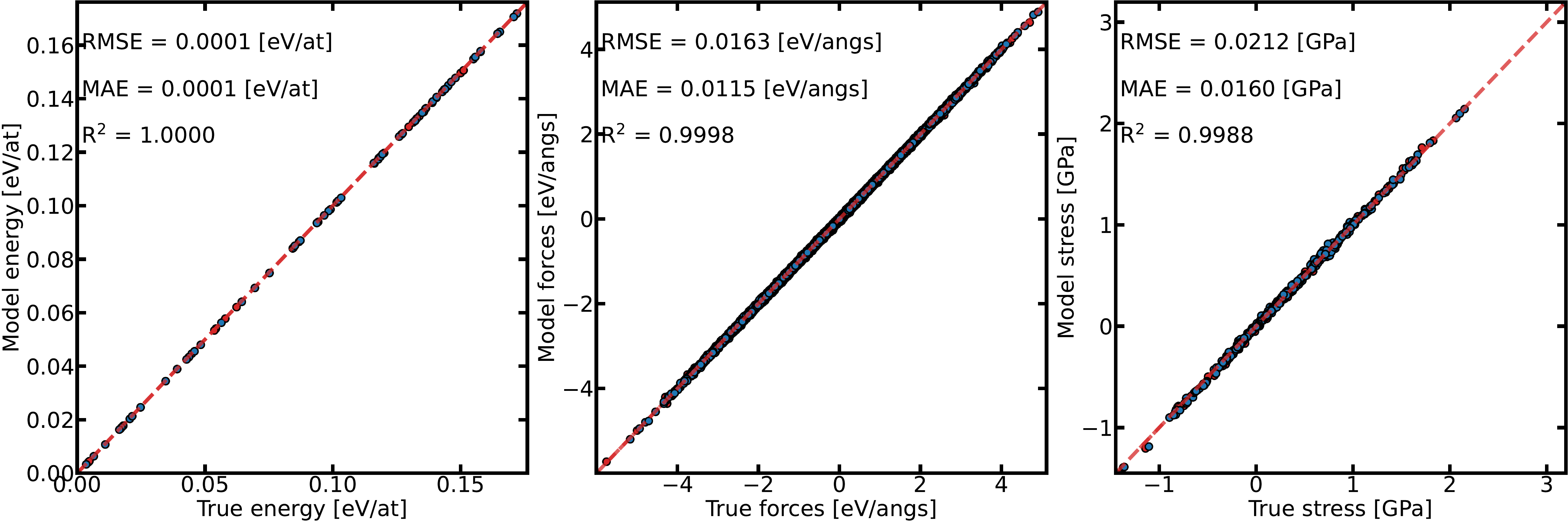}
    \caption{Correlation between energy, forces and stress computed with DFT and our MLIP.}
    \label{fig:mlip correlation}
\end{figure*}

\begin{figure}
    \centering
    \includegraphics[width=\linewidth]{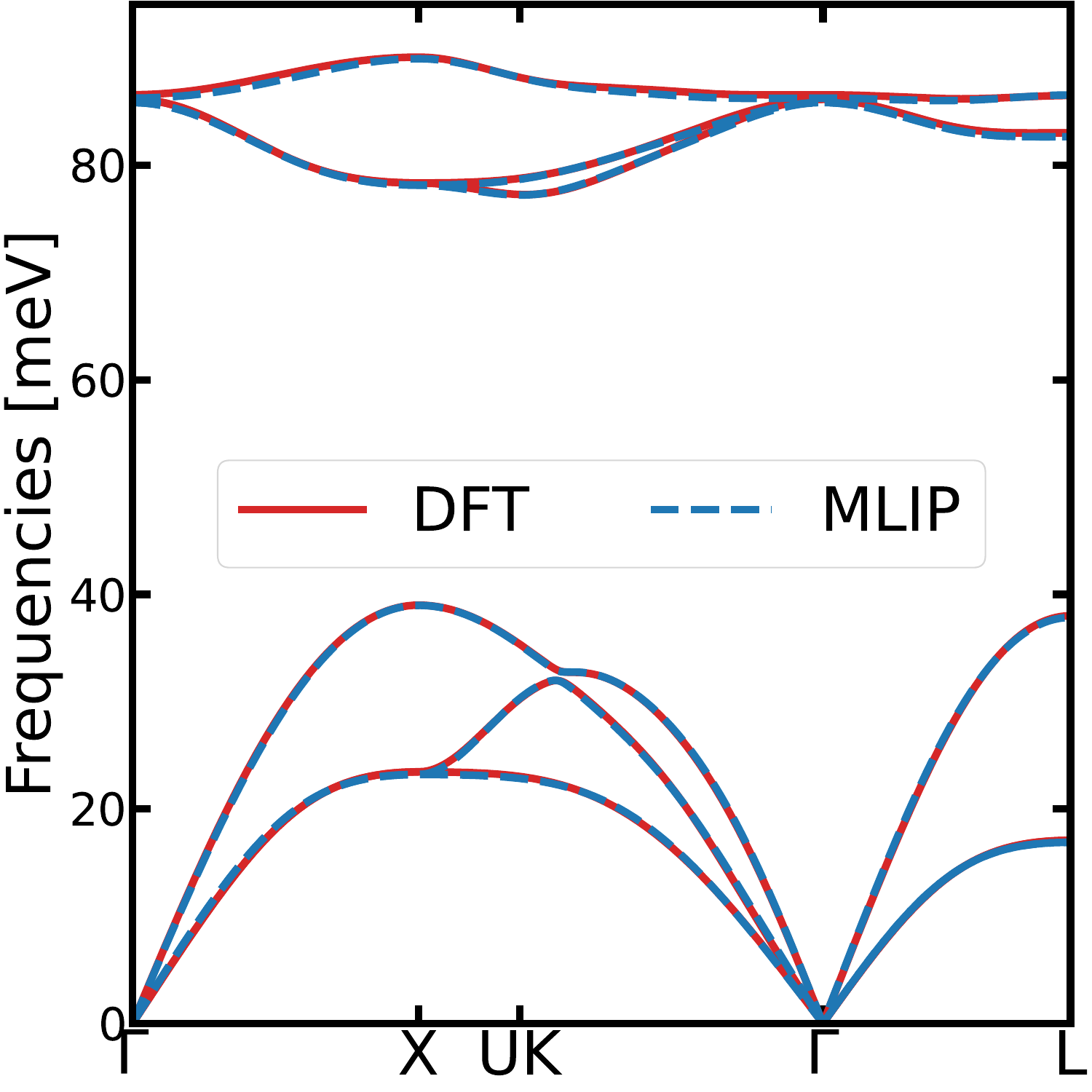}
    \caption{Comparison between harmonic phonons computed with finite difference using DFT and our MLIP.}
    \label{fig:harmonic comparison}
\end{figure}

\corrections{
\section{Validity of the Markovian approximation}

Our computational approach to incorporate the fluctuation-dissipation theorem is constructed on the premise that the Markovian approximation (ignoring memory effects) accurately models thermal transport.
Consequently, evaluating the correctness of this approximation is crucial.
To this end, we calculated the thermal conductivity without the Markovian approximation in the single mode approximation framework (ignoring collective effects) and without the fluctuation dissipation theorem.
Our non-Markovian expression for the thermal conductivity is given by
\begin{equation}
    \boldsymbol{\kappa}^{\text{nM}}_{\mathrm{SMA}} = \frac{\pi}{V} \sum_{\lambda\lambda'} \vec{v}_{\lambda\lambda'}\otimes \vec{v}_{\lambda\lambda'}\int_{-\infty}^{\infty} d\omega \chi_{\lambda}''(\omega)\chi_{\lambda'}''(\omega) c(\omega)
\end{equation}
where $c(\omega) = \omega^2 n(\omega) (n(\omega) + 1) / \kBT^2$ is the frequency dependent heat capacity, and $\chi_\lambda$ is the full susceptibility generalizing Eq.~\ref{eq:MarkovSusceptibility}.
When taking the Markovian limit, the thermal conductivity tensor is given by:
\begin{equation}
    \boldsymbol{\kappa}^{\text{M}}_{\mathrm{SMA}} = \frac{\pi}{V} \sum_{\lambda\lambda'} \vec{v}_{\lambda\lambda'}\otimes \vec{v}_{\lambda\lambda'} c(\Omega_{\lambda}) \Gamma_{\lambda\lambda'}
\end{equation}
where we define
\begin{equation}
    \Gamma_{\lambda\lambda'} = \frac{\Gamma_\lambda + \Gamma_{\lambda'}}{(\Omega_\lambda - \Omega_{\lambda'})^2 + (\Gamma_\lambda + \Gamma_{\lambda'})^2}.
\end{equation}
Focusing on the three-phonon scattering level, fig.\ref{fig:non-Markovian AgI} presents our findings for AgI, which is the most anharmonic material analyzed in this study.
\begin{figure}
    \centering
    \includegraphics[width=\linewidth]{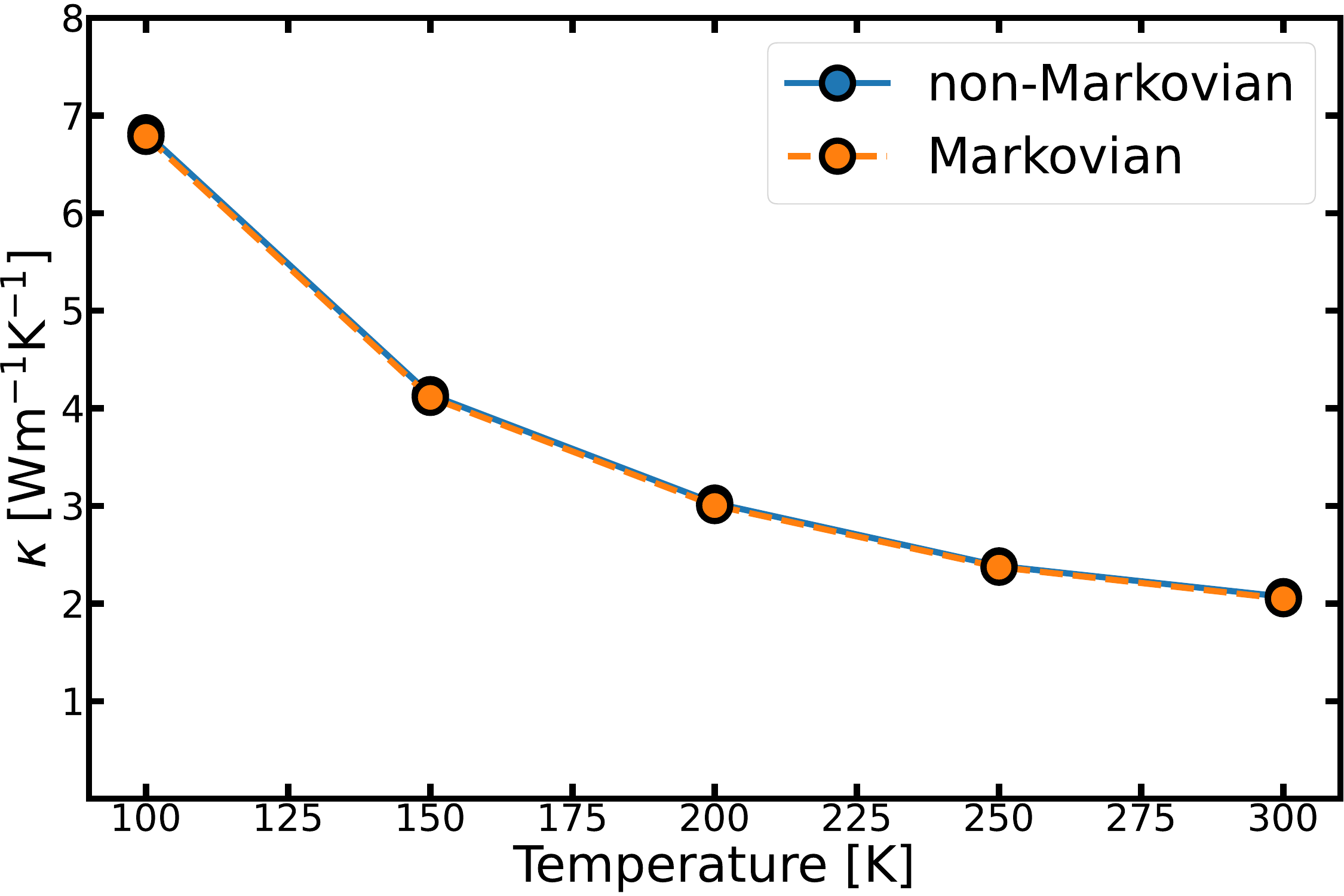}
    \caption{Comparison of thermal conductivity for AgI, with and without applying the Markovian approximation, utilizing three-phonon scattering and maintaining strict energy conservation.}
    \label{fig:non-Markovian AgI}
\end{figure}
The minor discrepancy between $\boldsymbol{\kappa}^{\text{M}}$ and $\boldsymbol{\kappa}^{\text{nM}}$ (about $10^{-2}$W/m/K) confirms the reliability of the Markovian approximation even in such highly anharmonic systems.
For BAs fig.\ref{fig:non-Markovian BAs} shows that the impact of non-Markovian effects is similarly minimal. 
Indeed, the relative difference between calculations with and without Markovian effects is around a few percent at the highest temperature, where memory effects are expected to be more significant.
\begin{figure}
    \centering
    \includegraphics[width=\linewidth]{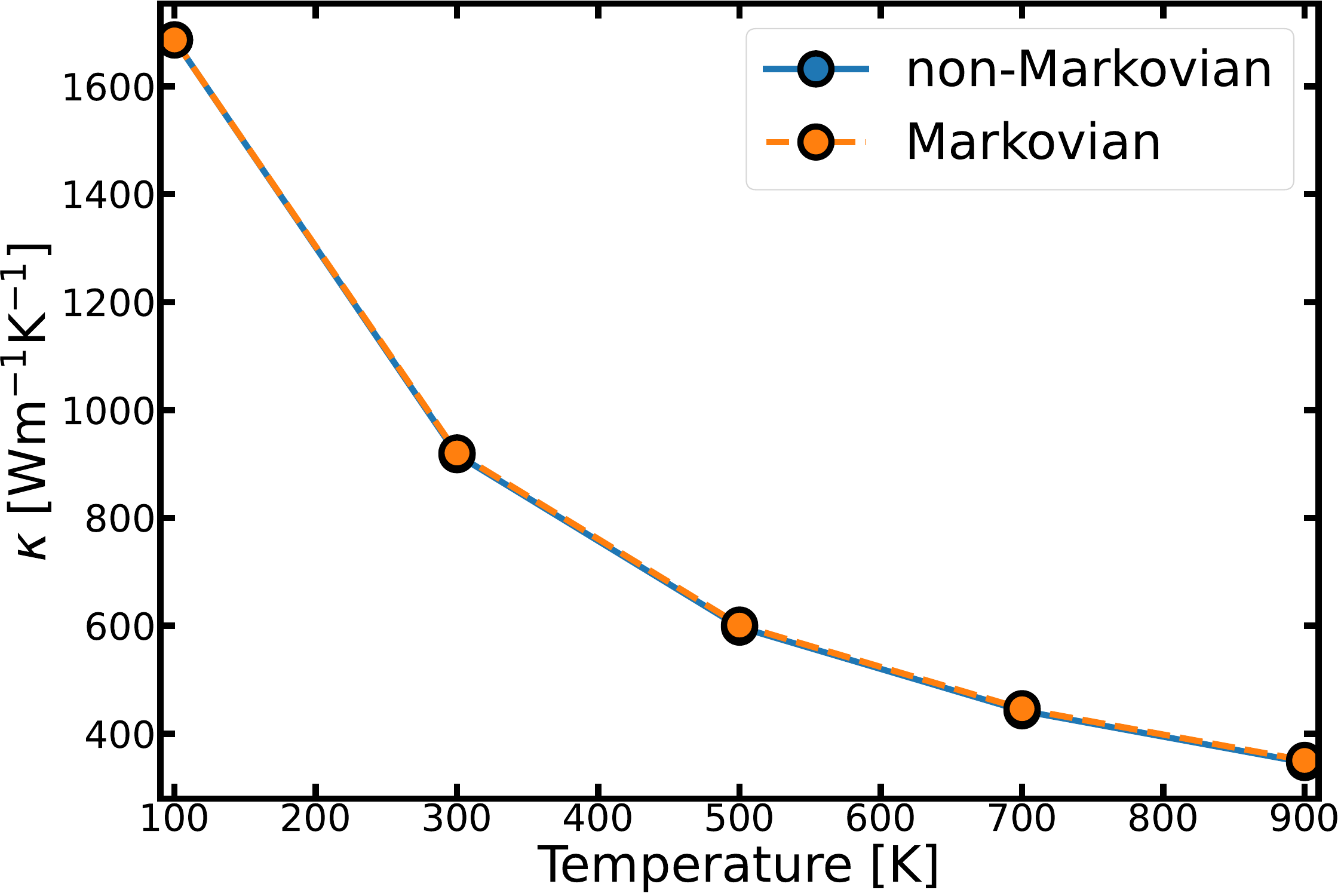}
    \caption{Comparison of thermal conductivity for BAs, with and without applying the Markovian approximation, utilizing three-phonon scattering and maintaining strict energy conservation. It is important to note that the collective effects, which are important for BAs, are not included in the calculations.}
    \label{fig:non-Markovian BAs}
\end{figure}

These tests are restricted to calculations with three-phonon interactions and omitted the collective phonon contributions.
The structure of the thermal conductivity formulation suggests that four-phonon interactions should not change the analysis performed at the three phonon level.
In particular, the Markovian limit of the thermal conductivity involves the substitution
\begin{equation}
    \int_{-\infty}^{\infty} d\omega c(\omega) \chi''_{\lambda}(\omega)\chi''_{\lambda}(\omega) \rightarrow \frac{c(\Omega_\lambda)}{2\Gamma_\lambda}.
\end{equation}
For modes that are strongly non-Markovian, the substituted formula involving the inverse of the scattering rates should be small, indicating a proportionally small contribution to the thermal conductivity.
%Thus, replacing this formula with the non-Markovian one should usually replace a negligible contribution with another negligible contribution, resulting in minimal difference of the integrated thermal conductivity tensor.
Furthermore, these contributions decrease even more when combined with group velocities, as non-Markovian modes are typically optical, and  have low group velocities.

For the collective contribution, there is currently no formulation to our knowledge involving non-Markovian effects. Such a derivation would be very interesting but we also expect these to be minimal in our case, since collective contributions are mainly observed in fairly harmonic systems where the Markovian limit is accurate, as in BAs.
For AgI, the collective effects in the Markovian approximation are even smaller than the memory contribution at the three-phonon level and we expect collective contributions to remain negligible also beyond the Markovian approximation.

We anticipate that incorporating four-phonon interactions and collective phonon contributions will not significantly alter the conclusions from three-phonon interactions in the single-mode approximation: the Markovian approximation remains applicable to the systems explored in this work.
}

\section{Computational details}

For the silicon and AgI results presented in this supplementary materials, we used the MLIP and parameters from reference \cite{Castellano2025}.
For BAs, all DFT calculations were performed with the Abinit suite~\cite{Gonze2020,Romero2020} using PBEsol pseudopotentials~\cite{Perdew2008} from the Pseudo-dojo project~\cite{vanSetten2018}.

After carefully checking the convergence, the kinetic energy cutoff was set to $42$~Hartree and a $10\times10\times10$ \textbf{k}-point grid was used for unit-cell calculations.
Supercell DFT calculations were performed on $3\times3\times3$ repetitions of the conventional unit-cells with only $\Gamma$ point sampling.

To reduce the computational cost, a machine learning interatomic potential (MLIP) was constructed using the Moment Tensor Potential framework~\cite{Shapeev2016,Novikov2021}.
The training dataset was generated self-consistently following the methodology described in ref~\cite{Castellano2025} and using the MLACS code~\cite{Castellano2022,Castellano2024b}, with temperatures generated randomly between $20$ and $1300$~K for the MLIP driven molecular dynamic simulations.
Only $122$ DFT configurations were necessary to converge the MLIP.
Our resulting potential provide a very accurate representation of the DFT described Born-Oppenheimer surface for BAs, as shown by the low root mean squared error and high correlation displayed on fig.(\ref{fig:mlip correlation}).
To further assess the accuracy of our MLIP, we also compare harmonic phonons computed with finite differences on fig.(\ref{fig:harmonic comparison}).
The MLIP and DFT dispersions are indistinguishable, giving a high confidence in the ability of our MLIP to describe the atomic motion in BAs crystals.

To compute the generalized interatomic force constants, NVT molecular dynamics simulations were performed on $4\times4\times4$ conventional supercells. The volume is obtained from averages of NPT MD runs, to account for the system's thermal expansion.
All the MD calculation were performed with the LAMMPS package~\cite{Thompson2022}, and the mode-coupling theory calculation were done with the TDEP package~\cite{Hellman2011,Hellman2013a,Hellman2013b,Knoop2024}.
For BAs, the generalized interatomic force constants used cutoffs of $9.55$, $6.0$ and $3.5$~$\mathring{A}$ for the second, third and fourth-order, respectively.
For the thermal conductivity calculations, we used a full \textbf{q}-point grid of $32\times32\times32$ and the scattering were computed on $16\times16\times16$ and $6\times6\times6$ Monte-Carlo \textbf{q}-grid for the third and fourth order, using the implementation presented in ref~\cite{Castellano2025}.

\begin{figure}
    \centering
    \includegraphics[width=\linewidth]{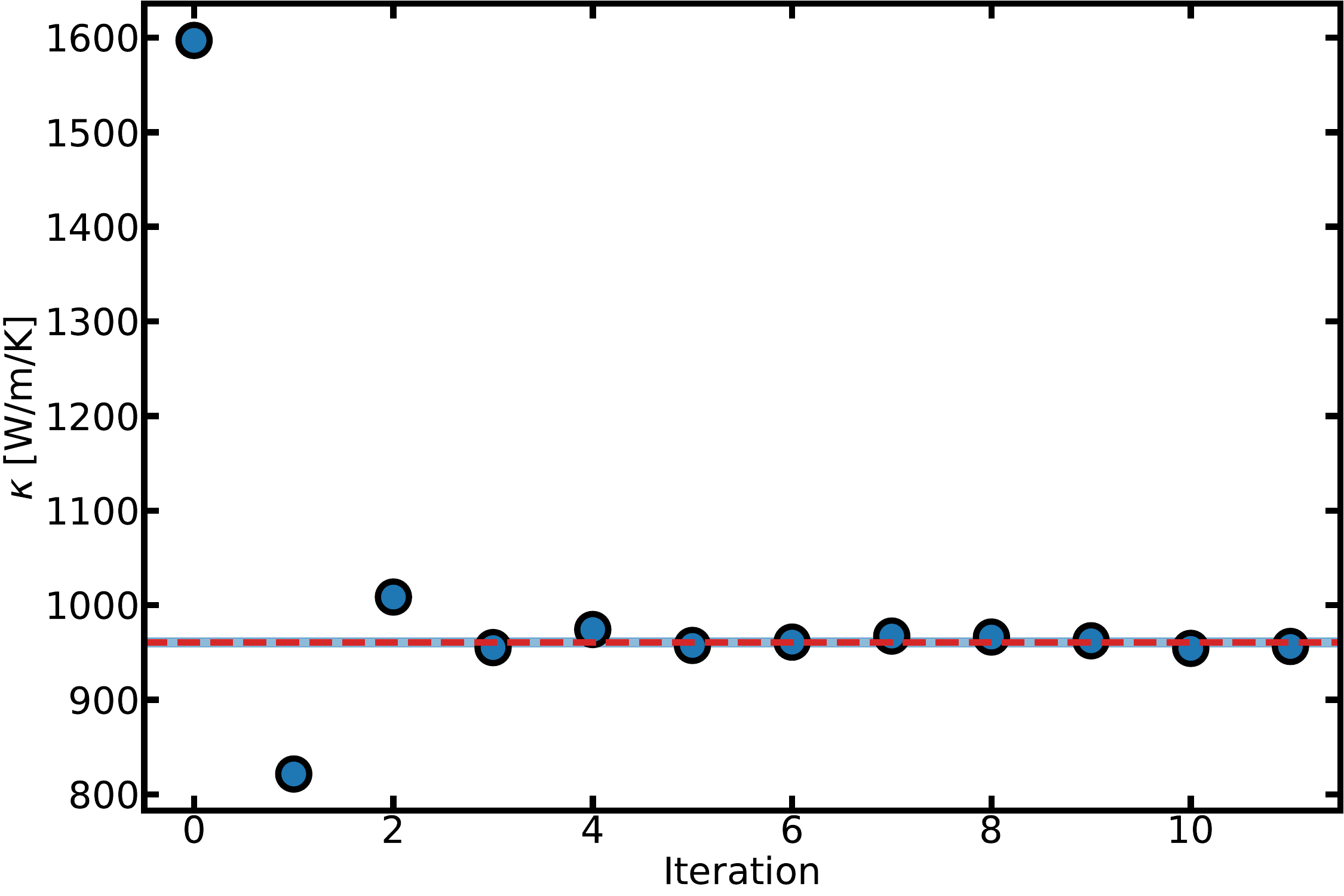}
    \caption{Convergence of the thermal conductivity in BAs with FD iterations. The calculation only consider three-phonon interactions.}
    \label{fig:ConvergenceBAs_300K}
\end{figure}

\corrections{Fundamentally, the self-consistent algorithm involves iteratively performing thermal conductivity calculations using Lorentzians whose broadening is derived from previous iterations.
In all instances, the initial iteration employs energy conservation within the adaptive Gaussian scheme~\cite{Castellano2025}, although it has been observed that the starting point does not affect the final results.
No mixing of the linewidths between iterations was used, as this approach did not enhance convergence.
For all calculations, we conducted 10 iterations and averaged the results over the last 5 iterations, to help mitigate noise arising from the Monte-Carlo grid integration.
In both the main and supplementary texts, all results incorporating the fluctuation-dissipation condition also display the variance of the thermal conductivity over these 5 iterations as a shaded area. However, the variance is usually smaller than the width of the lines.
The self-consistent cycle is illustrated in Fig.\ref{fig:ConvergenceBAs_300K}, which depicts the evolution of the thermal conductivity with iterations for BAs at 300K at the three-phonon level.}

\bibliography{biblio}% Produces the bibliography via BibTeX.